\documentclass[11pt]{article}
\usepackage{amsmath}
\usepackage{amsfonts}
\usepackage{graphicx}
\usepackage{authblk}
\usepackage{esvect}

\usepackage{ulem}
\usepackage{color}

\newcommand{\bv}[1]{\boldsymbol{#1}}

\newcommand{\qu}[1]{``#1''}

\newcommand{\K}{\bv K}

\newcommand{\U}{\bv U}
\newcommand{\V}{\bv V}
\newcommand{\X}{\bv X}
\renewcommand{\u}{\bv u}
\renewcommand{\v}{\bv v}

\newcommand{\C}{\bv{C}}
\newcommand{\x}{\bv{x}}

\newcommand{\oneoversqrt}[1]{\frac{1}{\sqrt{#1}}}
\newcommand{\phiX}{\Phi\parens{\X}}
\newcommand{\phix}{\Phi\parens{\x}}

\newcommand{\parens}[1]{\left(#1\right)}

\renewcommand{\exp}[1]{\text{exp}\parens{#1}}
\newcommand{\bracks}[1]{\left[#1\right]}
\newcommand{\threevec}[3]{\bracks{\begin{array}{c} #1 \\ #2 \\ #3 \end{array}}}
\newcommand{\threebythreemat}[9]{\bracks{\begin{array}{ccc} #1 & #2 & #3 \\ #4 & #5 & #6 \\ #7 & #8 & #9 \end{array}}}
\newcommand{\expe}[1]{\mathbb{E}\bracks{#1}}
\newcommand{\cexpe}[2]{\expe{#1\,|\,#2}}

\newcommand{\beqn}{\vspace{-0.25cm}\begin{eqnarray*}}
\newcommand{\eeqn}{\end{eqnarray*}}
\newcommand{\bneqn}{\vspace{-0.25cm}\begin{eqnarray}}
\newcommand{\eneqn}{\end{eqnarray}}

\definecolor{black}{rgb}{0,0,0}
\definecolor{blue}{rgb}{0,0,0.7}

\definecolor{green}{rgb}{0.133,0.545,0.133}

\definecolor{yellow}{rgb}{1,0.549,0}

\definecolor{red}{rgb}{1,0.133,0.133}

\definecolor{purple}{rgb}{0.58,0,0.827}

\definecolor{brown}{rgb}{0.55,0.27,0.07}

\begin{document}

\title{Using Regression Kernels to Forecast A Failure to Appear in Court}
\author[1,2]{Richard Berk\thanks{Electronic address: \texttt{berkr@sas.upenn.edu}; Corresponding author}}
\author[2]{Justin Bleich}
\author[2]{Adam Kapelner}
\author[3]{Jaime Henderson}
\author[3]{Geoffrey Barnes}
\author[3]{Ellen Kurtz}
\affil[1]{Department of Criminology, University of Pennsylvania}
\affil[2]{Department of Statistics, University of Pennsylvania}
\affil[3]{Philadelphia Adult Department of Probation and Parole}
\maketitle

\begin{abstract}
Forecasts of prospective criminal behavior have long been an important feature of many criminal justice decisions. There is now substantial evidence that machine learning procedures will classify and forecast at least as well, and typically better, than logistic regression, which has to date dominated conventional practice. However, machine learning procedures are adaptive. They ``learn'' inductively from training data. As a result, they typically perform best with very large datasets. There is a need, therefore, for forecasting procedures with the promise of machine learning that will perform well with small to moderately-sized datasets. Kernel methods provide precisely that promise. In this paper, we offer an overview of kernel methods in regression settings and compare such a method, regularized with principle components, to stepwise logistic regression. We apply both to a timely and important criminal justice concern: a failure to appear (FTA) at court proceedings following an arraignment. A forecast of an FTA can be an important factor is a judge's decision to release a defendant while awaiting trial and can influence the conditions imposed on that release. Forecasting accuracy matters, and our kernel approach forecasts far more accurately than stepwise logistic regression. The methods developed here are implemented in the \texttt{R} package \texttt{kernReg} currently available on \texttt{CRAN}.
\end{abstract}

\section{Introduction}

Conventional criminal justice risk assessments have their roots in parole decision-making dating back to the 1920s (Borden, 1928). The roots run deep. Even very recent methods typically rely on scaling approaches developed by Earnest Burgess shortly after World War I (Burgess, 1928). 

Criminal justice risk assessment technology begins with one or more behavioral outcomes to be forecasted. Felony arrests are an example. A search for ``risk factors'' associated with the outcomes typically follows. Prior record, age and gender are among the risk factors usually found. The risk factors are ordinarily combined in a linear fashion to produce a numerical score. The higher the score, the greater the presumptive risk. Subsequently, when measures of the risk factors exist, but a behavioral outcome is not yet known, a risk score can be computed and used to make a forecast. The forecast either can be a score that attempts to capture the degree of risk or can be translated into a binary outcome class such as ``fail'' or ``not fail'' using a threshold on that score. Scores above the threshold forecast one class. Scores at or below the threshold forecast the other class. Although it is often difficult to determine how accurate such forecasts are (Reiss, 1951; Farrington and Tarling, 2003; Gottfredson and Moriarty, 2006; Ridgeway, 2013b), they are now routinely used to inform a variety of criminal justice decisions (Berk and Bleich, 2013). 

Because the operational outcomes are typically binary, most of the risk assessment instruments developed over the past several decades have used logistic regression to determine the relevant risk factors. There is now strong evidence that more accurate forecasts can be obtained directly from machine learning procedures (Berk, 2012; Berk and Bleich, 2013; Ridgeway, 2013; Bushway, 2013) such as random forests (Breiman, 2001), stochastic gradient boosting (Friedman, 2002), Bayesian additive regression trees (Chipman et al., 2010) and support vector machines (Vapnick, 1998). These are ``black box'' algorithms able to construct complicated ``profiles'' that sort individuals into discrete outcome classes such as an arrest for one of several different kinds of crime. There is no need for explicit identification of risk factors if accurate forecasts are the primary goal.

Machine learning procedures are based on adaptive methods that work best on very large datasets compatible with extensive data exploration. For many criminal justice applications, such datasets exist and are reasonably accessible. But there are also criminal justice settings for which ``small data'' are more appropriate or are all that can be obtained. For example, there are now a variety of highly specialized courts that handle only firearm-related crimes or only drug crimes or only sex crimes. Several hundred cases may be heard each year rather than tens of thousands. There is also an enormous number of small jurisdictions from which ``big data'' are unlikely. We need, therefore, forecasting procedures having many of the performance benefits of machine learning without relying on a large number of observations.

Recent extensions of logistic regression that work well in samples of modest size would seem to qualify. Regressors are ``kernel'' transformations of the original predictors that will often include \textit{a priori} the same kinds of subtle profiles inductively discovered by machine learning. In addition, a variety of methods can be applied that down-weight features of the kernel having weak associations with the response variable. The combination of kernel transformations and down-weighting allows for the number of predictors to be as large or larger than the number of observations. It costs only additional computer time to introduce far more potential predictors than conventional regression can possibly manage. 

In this paper, we describe and apply kernel regression procedures to develop forecasts of a common binary outcome following an arraignment: a failure to return to court after a release awaiting trial. At least since the Manhattan Bail Project in 1961, there have been serious efforts to reform the ways bail decisions are made (McElroy, 2011). Among the most important changes have been to take quantitative risk assessments far more seriously. Burgess-like scales are used to inform bail decisions and procedural reforms (Clarke et al., 1976; Goldkamp and White, 2006; VanNostrand and Keebler, 2009; Bornstein et al., 2012; Arnold Foundation, 2013). For example, a decision by a judge at an arraignment to release a defendant while the defendant is awaiting trial can be substantially influenced by a forecast of ``future dangerousness'' or a forecast that an individual will not return to court when later required to do so. Clearly, a lot is at stake. Defendants may be held when there is no need or defendants may be released when some form of incapacitation is required. Forecasting accuracy really matters. 

Section 2 provides a didactic discussion of kernels that can be used in regression analysis. After a gentle introduction to kernel transformations in Section 2.1, the rest of Section 2 provides a summary of the underlying statistical theory, a rationale for kernel dimension reduction, a discussion of asymmetric costs from forecasting errors, an explanation of tuning, and a proper context for statistical inference. We also briefly discuss the implementation of our procedure, which is found in the \texttt{R} package \texttt{kernReg} currently available on \texttt{CRAN}. Section 3 is devoted to forecasting whether after an arraignment defendants later return to court when required to appear. The forecasting task is very challenging because of important omitted variables and little convincing theory to guide model specification. Results from conventional stepwise logistic regression and our proposed ``regularized'' kernel logistic regression are compared. The legitimacy of such comparisons is discussed as well. In Section 4, we conclude and provide recommendations to the practitioner who wishes to employ our methods. 

An important feature of the paper is an attempt to demystify kernel regression procedures. There are many demanding details, some of which may be unfamiliar to social scientists. Readers interested primarily in the empirical results may wish to skip to Section 3 after reading through Section 2.1. For those readers, kernel regression can be summarized as a form of nonparametric regression in which an extremely rich menu of predictor transformations is used to fit very complex relationships. In the spirit of machine learning, kernel regression is a ``black box''  method. The ways in which predictors are related to the response are not readily apparent. Thus, we are inclined to view kernel regression as a form of machine learning for ``small data.''

\section{Kernel Regression Methods for Forecasting}

In a conventional regression analysis, a primary goal is to represent how one or more predictors are related to a response. Often those relations are interpreted as causal. But there can also be interest in the fitted values. Sometimes the fitted values are plotted to provide information about the possible nonlinear functional forms. There may be no regression coefficients to interpret, but the intent is still to characterize how the predictors are related to the response. Partial response plots used with generalized additive models are a good illustration (Hastie and Tibshirani, 1990).

Sometimes the fitted values by themselves are of interest. For example, when the response is categorical, the fitted values can be used for classification. The goal might be to determine whether particular transactions are fraudulent. Or the goal might be to provide a diagnosis for patients exhibiting certain symptoms. Such goals do not require that the relationships between the predictors and the response be captured in ways that are substantively interpretable. For instance, in principle components regression, the regressors are linear combinations from the full set of original predictors. Although post hoc interpretative overlays are sometimes employed, how the predictors are related to the response is typically obscured. The fitted values are the essential motivator.

Forecasting is another activity in which the role of predictors need not be a primary concern. An investor might be deciding which energy futures are a good bet based on their forecasted returns a year hence. Or a parole board may decide which inmates to release based on forecasts of whether a violent crime will be committed. One may achieve excellent forecasting accuracy with no real understanding about how the predictors are related to the response (Berk, 2012; Berk and Bleich, 2013; Ridgeway, 2013b, Bushway, 2013). Indeed, it is often productive to make forecasting and explanation separate data analysis objectives.

If the focus can be exclusively on forecasting, one has the opportunity to employ predictors in a manner that may dramatically improve forecasting accuracy even if explanation is severely compromised. Machine learning is a set procedures that commonly makes this tradeoff; kernel methods do likewise. 

Here, we focus on kernel methods for regression applications in which the primary interest is in fitted values and subsequent forecasts. We will see that when complicated nonlinear and/or interaction effects may be needed, but the precise functions are unknown or the relevant variables are not in the data set, kernel methods can automatically assemble a very rich menu of functions that may serve as an effective alternative. Forecasts of useful accuracy can follow. Although we will later emphasize regression with binary outcomes, kernel methods can in principle be used in any form of regression when one is trying to characterize the distribution of a response variable conditional on a set of predictors.
 
\subsection{Linear Basis Expansions} 

Linear basis expansions can be building blocks for a wide variety of statistical procedures and are a critical starting point for a discussion of how kernels can be employed in regression (Hastie et al., 2009: Section 5.1). For a single predictor $X$, 

\begin{equation}
f(X) = \sum_{m=1}^{q} \beta_{m}\phi_{m}(X),
\label{eq:basis}
\end{equation}
where the predictor $X$ is replaced by a sum of $q$ transformations of $X$, each transformation represented by $\phi_{m}(X)$. A cubic function in X is a simple illustration: {$\phi_1(X) = X,~\phi_2(X) = X^2$ and $\phi_3(X) = X^3$}. Here, $q = 3$, and the three transformations of $X$ are $X, X^{2},$ and $X^{3}$; $\phi_1(X) = X$ is a ``trivial transformation.'' The corresponding weights $\beta_{1}, \beta_{2}$, and $\beta_{3}$ can be conventional regression coefficients. Where there was initially a single function of $X$, there are now three functions of $X$: hence the term ``expansion." Other kinds of linear basis expansions include trigonometric functions, indicators variables, and various types of splines (Haste et al., 2009, Chapter 5). The formulation also is readily extended to more than one predictor. When linear basis expansions are used kernel applications, it is common to used the notation $\Phi(\textbf{X})$ to represent the collection of linear basis expansions for the full set of predictors. 

\begin{figure}[htp]
\begin{centering}
\includegraphics[width=5in]{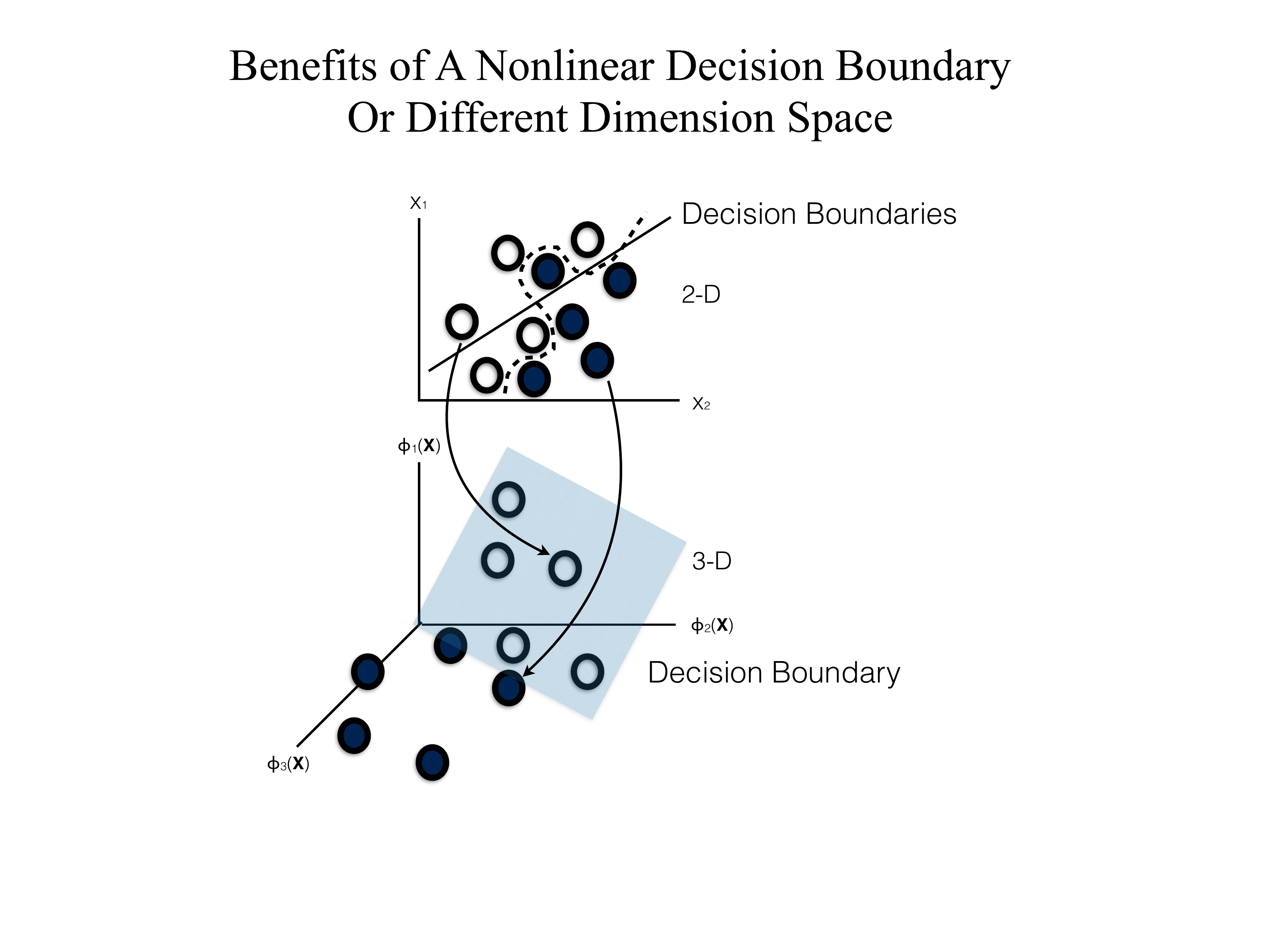}
\caption{An Illustration of the Gains From a Nonlinear Fit (top figure) or a Transformed Predictor Space (bottom figure)} 
\label{fig:d}
\end{centering}
\end{figure}

The benefits from linear basis expansions depend on two potential consequences of Equation~\ref{eq:basis}. First, the expansion can directly alter how the relationships between the response and the predictors are characterized. In the example just given, a cubic function may fit the data better. Second, by transforming the \textit{space} in which the observations are located, patterns may be found that are otherwise obscure. Precisely how this can be done is considered in later sections. For now, Figure~\ref{fig:d} illustrates both possibilities.

The upper part of Figure~\ref{fig:d} shows a scatter plot with two predictors, $x_{1}$ and $x_{2}$. For example, $x_{1}$ could be the number of prior arrests, and $x_2$ could be age.\footnote
{
For ease of exposition, we are playing a little fast and loose with notation at this point because formally both predictors are vectors. We will get more formal shortly. 
}

The open circles represent one of two response outcomes (e.g., rearrested while on parole). The solid circles represent the other response outcome (e.g., not rearrested while on parole). There are also two overlays representing two different decision boundaries. The solid line is a linear decision boundary. The dashed line is a nonlinear decision boundary.\footnote
{
The term ``decision boundary'' is used because depending on which side of the decision boundary an observation falls, different decisions about that observations can be justified. For example, one decision might be to release an inmate on parole and another decision might be to be keep the inmate incarcerated. 
} 

The data analyst's task is to partition the predictor space using a decision boundary so that all of the open circles fall in one partition and all of the solid circles fall in the other partition. If one were able to do so, the pair of predictors $x_1$ and $x_2$ would be able to perfectly distinguish between the two outcomes. The predictors would be able to classify these observations without error. 

It is impossible here to find a linear decision boundary in the 2-D predictor space that perfectly distinguishes between the open and solid circles. Any linear attempt to classify cases in these two dimensions will result in two partitions of the space, with at least one having a mix of open and solid circles. More technically phrased, there can be no linear ``separation" between the open and solid circles. In Figure~\ref{fig:d}, for instance, one solid circle falls in the partition dominated by open circles. 

However, Figure~\ref{fig:d} shows a nonlinear decision boundary that can partition the 2-D predictor space discriminating the two response types perfectly. Nonlinear transformations of the predictors can in principle be helpful in precisely this way.\footnote
{
There are two ways to think about this. In the original units of the predictors, the decision boundary is nonlinear. Or in the units of the transformed predictors, the decision boundary is linear. We show the former in Figure~\ref{fig:d}.
} 

The lower part of Figure~\ref{fig:d} has three predictors, $\Phi_{1}(\textbf{X}),~ \Phi_{2}(\textbf{X}),$ and $\Phi_{3}(\textbf{X})$, where $\textbf{X}$ is matrix notation representing both $x_{1}$ and $x_{2}$. Thus each predictor is a different basis expansion term using \textit{both} $x_{1}$ and $x_{2}$. For example, the expansion might be a cubic polynomial of an element by element product $x_{1} \times x_{2}$, where $q=3$.\footnote
{
``Element by element'' means $x_{11} \times x_{12}, ~ x_{21} \times x_{22} ,~ \dots ~, x_{N1} \times x_{N2}$, where the first subscript denotes the observation number and the second subscript is for the predictor number. $\Phi_{1}(\textbf{X})$ is then the element by element product, $\Phi_{2}(\textbf{X})$ is the element by element product squared, and $\Phi_{3}(\textbf{X})$ is the element by element product cubed. One might view the three terms as an interaction effect variable, an interaction effect variable squared, and an interaction effect variable cubed.
}

Each dimension represents a term of the expansion that as a group defines a 3-D space in which the observations can be located. In this new space, one can see that the open circles are separated perfectly from the solid circles because the former are located toward the back of the figure, and the latter are located toward the front of the figure. Therefore, it is possible to construct a 2-D plane that can perfectly discriminate between the two response types. 

Linear basis expansions are easily extended to higher dimensions. Figure~\ref{fig:d} provides an initial sense of the benefits that we will addressed in more depth later. But in practice, perfect separation is still very difficult to achieve. Rather, we seek substantially improved separation. 

\subsection{Kernel Functions and Kernel Matrices}

A powerful way to construct and deploy linear basis expansions is to apply ``kernel transformations.'' Kernel transformations are defined by a ``kernel functions.'' There are many such functions. Some are typically employed in highly specialized applications. Still, this coupling of kernel to application is usually justified by little more than hunch or craft lore (Gross et al., 2012; Duvenaud et al., 2013). We consider here two kernel functions commonly used in regression settings that seem to work well.

We begin with a toy predictor matrix $\textbf{X}$: 

\begin{equation}
\textbf{X}=
\left[ \begin{array}{ccccc}
1 & 2 & 3 & 2 & 0 \\
2 & 6 & 1 & 1 & 1 \\
0 & 6 & 0 & 1 & 2 \\
\end{array} \right]
\end{equation}
There are 3 rows representing 3 observations, where the number of observations is conventionally denoted by $N$. There are 5 columns representing 5 predictors, where the number of predictors is conventionally denoted by $p$. To illustrate some important features of kernels, there are more predictors than cases (i.e., $p > N$). This does not present an immediate problem but would if we considered off-the-shelf regression tools. 

Because of the nature of the kernel functions to be applied, all of the elements in \textbf{X} must be numerical. This includes categorical variables. If $C$ is the number of categories, it is conventional to use $C-1$ indicator variables, all coded numerically in the same way (e.g., 0 or 1). For example, if there are four different kinds of employment (including not being employed), there would be three indicator variables, where for each, ``1'' represents the presence of that form of employment and ``0" represents the absence of that form of employment. This is consistent with common practice in many different kinds regression applications. 

\subsubsection{The Radial Basis Kernel}

It is nearly universal to denote a kernel function by $k(\x, \x')$ where $\x$ and $\x'$ are two different row vectors in $\textbf{X}.$\footnote
{
In standard notation, the two row vectors are in $\mathbb{R}^p$, which is a $p$-dimensional Euclidian space. Here, $p =5$ and a row vector is for a given observation its value for each predictor. For example, age might be 24, years since the last arrest might be 2.5, the number of prior prison terms might be 2, gender might be male (i.e. ``1''), and the number of prior convictions might be 3. 
}
The \textit{radial basis kernel} is defined as 
\begin{equation}
k(\x,\x') = \exp{-\gamma \| \x - \x{'} \|^2},
\label{eq:rad}
\end{equation} 
with $\|.\|$ denoting the squared Euclidian distance (i.e. the ``sum of squared differences'' also, known as the ``norm''), and $\gamma$ denoting a scale parameter greater than 0.

The kernel transformation is created by applying the kernel function to the data $\X$ producing the \textit{kernel matrix,} a matrix which, as we will see shortly, contains the predictive information for a proper regression analysis. To arrive at the \textit{kernel matrix} $\textbf{K}$, one computes $k$ for each combination of rows $i, j$ and inserts the kernel value in the $i,j$ location of $\textbf{K}$. Since there are $N$ observations, there are $N$ comparisons for each observation yielding an $N \times N$ matrix. As an example, consider the second and third row of our toy $\textbf{X}$. One has for the sum of squared differences: $ (2-0)^2 + (6-6)^2 + (1-0)^2 + (1-1)^2 + (1-2)^2= 6$. The sum of squared differences is multiplied by scale parameter $\gamma$, negated, and then exponentiated. If the scale parameter were $0.01$, one can perform all $3 \times 3 = 9$ calculations to compute

\begin{equation}
\textbf{K} =
\left[ \begin{array}{ccc}
1.0 & .79 & .73 \\
.79 & 1.0 & .94 \\
.73& .94 & 1.0 \\
\end{array} \right].
\end{equation}
All kernels matrices are symmetric. Element $i,j$ is the same as element $j,i$.

The diagonal entries of the radial basis kernel are always 1 (because the squared distance between any $\textbf{x}$ and itself is 0 and $\exp{0} = 1$), and the off-diagonal entries are between 0 and 1 (because squared distances are positive and $\exp{-\Delta d^2}$ is bounded between 0 and 1 for positive $\Delta d^2$).\footnote
{
Thus there are only $\binom{N}{2} - N$ computations to construct $\K$.
} 
Radial kernels and others that build on Euclidian distances yield $\textbf{K}$'s that can be viewed as similarity matrices. Because of the $-\gamma$ in Equation~\ref{eq:rad}, larger off-diagonal values imply less distance between a given pair of observations, which means that they have more similar profiles over variables --- they are more similar. Radial basis kernels have proved to be useful in a wide variety of applications but for regression, there can be a better choice.

\subsubsection{The ANOVA Radial Basis Kernel}

The ANOVA radial basis kernel is closely related to the radial basis kernel. Using common notation for the ANOVA kernel,
\begin{equation}
k(\x,\x{'})= \left ( \sum_{j=1}^{p} \exp{ -\gamma (x_{j} - x'_{j})^{2}} \right )^{d},
\end{equation}
where $x_{j}$ and $x'_{j}$ are two different observations' values for predictor $j$, and $p$ is the number of predictors in $\textbf{X}$.\footnote
{
The computational translation is a little tricky. Here are the steps to compute $\textbf{K}$: (1) for observations $i$ and $j$ do an element by element subtraction over each of the predictors; (2) square each of the differences; (3) multiply each of these squared differences by minus $\gamma$; (4) exponentiate each of these products; (5) sum the exponentiated products; (6) raise the sum to the power of $d$; and (7) Repeat steps 1-6 for all pairs of observations $i,j$ to compute all $N \times N$ entries in $\K$.
}
Because the computations begin with differences, which after being transformed are added together, the calculations are linear when $d=1$, and one has a linear (additive) similarity formulation. When $d=2$, one has a formulation with products that can be seen as two-way interactions and a squared similarity formulation. By the same reasoning, when $d=3$, one has three-way interactions and a cubic similarity formulation.\footnote
{
To take a simple example, suppose there are three predictors. For the pair of observations from the first and second row of $\textbf{X}$ with $\gamma = 1$ and $d=1$, the sum of differences is $\exp{-(x_{11} - x_{12})^2} + \exp{-(x_{12} - x_{22})^2} + \exp{-(x_{13} - x_{23})^2}$. This is linear and additive. For $d=2$, the result is $[\exp{-(x_{11} - x_{12})^2} + \exp{-(x_{12} - x_{22})^2} + \exp{-(x_{13} - x_{23})^2}]^{2}$. All of the terms are now products of two variables, which are two-way interaction effects. For $d=3$, the result is $[\exp{-(x_{11} - x_{12})^2} + \exp{-(x_{12} - x_{22})^2} + \exp{-(x_{13} - x_{23})^2}]^{3}$. All of the terms are now products of three variables, which are three-way interaction effects. 
} 
The result here for $\gamma =.01$, and $d = 2$ is

\begin{equation} 
\label{eq:anovaK}
\textbf{K} =
\left[\begin{array}{ccc}
25.00 & 22.88 & 12.16 \\
22.88 & 25.00 & 24.41 \\
22.16 & 24.41 & 25.00 \\
\end{array} \right].
\end{equation}

The value of $d$ is commonly set at 1, 2, or 3. In our experience, using 2 or 3 seems to work well in practice. The values for $\gamma$ are generally far more important and much more difficult to determine. With larger values of $\gamma$, the off-diagonal values of $\textbf{K}$ become smaller.  Their \textit{different} Euclidian distances are reduced. In a regression setting, this will make the support from transformed predictor matrix more localized so that more complex relationships with the response variable can be captured. In the language of smoothers, a smaller window (or band width) is being used. 

\begin{figure}[htp]
\begin{centering}
\includegraphics[width=5in]{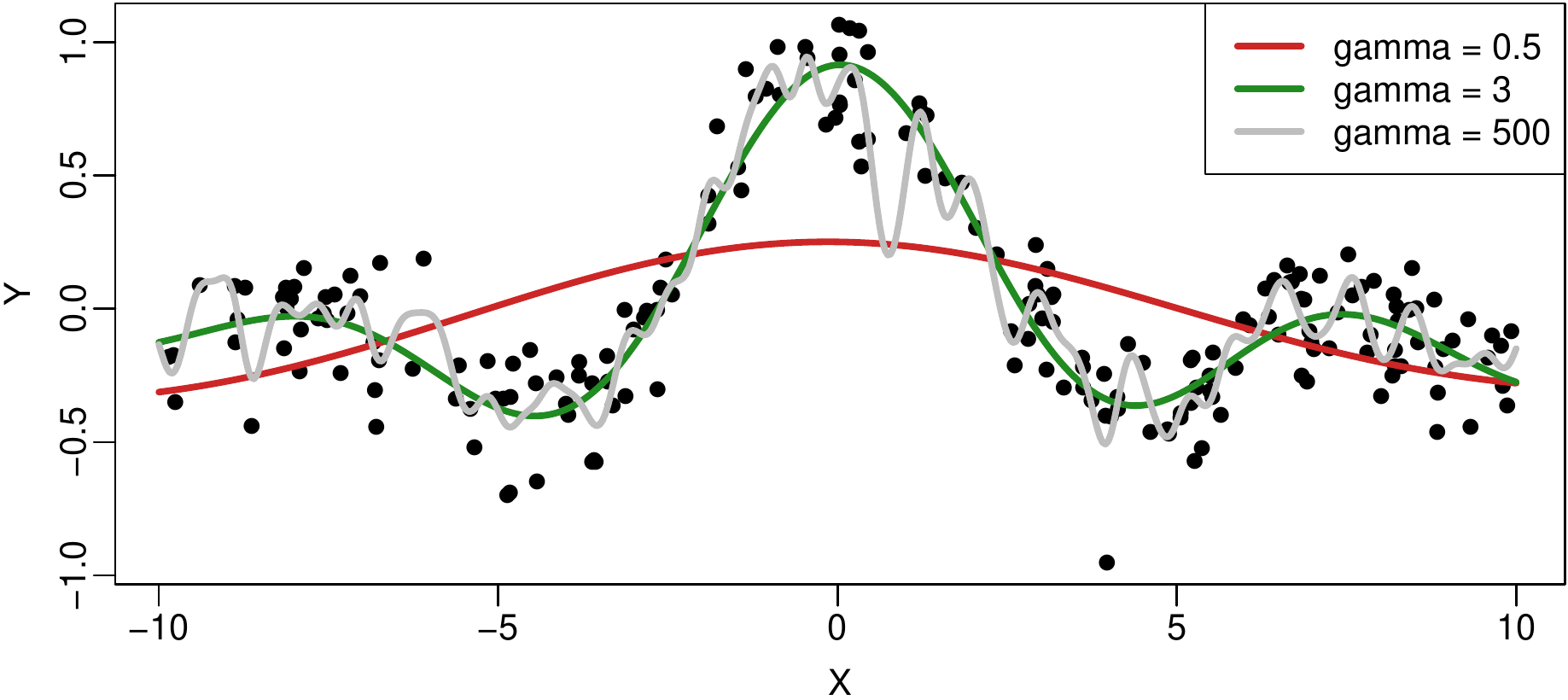}
\caption{A Simulation of How Fitted Values Depend on the Value of $\gamma$} 
\label{fig:sim}
\end{centering}
\end{figure}

Figure~\ref{fig:sim} illustrates these points. The figure is a conventional scatterplot showing the results of a simulation in which the fitted values from a simple kernel regression depend on the value of $\gamma.$ To help make the plot more visually instructive, both the response and single predictor are quantitative. For the same reason, we use a rectangular predictor distribution so there is no significant data sparsity, and the response is highly nonlinear function of the predictor. 

In this simulation, $d=1$ because there is only one regressor (no interactions are possible), and the fit is quite good when $\gamma = 3$. For a $\gamma$ of 0.5, the fitted values are far too smooth. Important patterns are not captured, although there is much less variance to contend with. For a $\gamma$ of 500, the fitted values are much too rough. Patterns are captured that are dominated by noise. 

There is a lot going on beneath the surface. The $\textbf{K}$ constructed for the simulation is not a conventional covariance matrix nor a conventional smoother matrix, and nowhere are the predictors in $\textbf{X}$ or the linear basis expansions $\Phi(\textbf{X})$ explicitly represented. We will see that this all makes sense because of the ``kernel trick'' in which $\K = \phiX\phiX^\top$. In the next several pages, we summarize the reasoning.

\subsection{How the Kernel Works for Regression}

Broadly stated, the operational procedures for the kernel-based forecasting procedures we apply  are relatively straightforward. The set of predictors is transformed using an ANOVA kernel. Principle components analysis is applied to the kernel. Then, logistic regression is implemented using a subset of the principle components as regressors. The process of kernel construction, principle component analysis and logistic regression is repeated a number of times with different tuning parameters for the kernel and different subsets of principle components. Using out-of-sample performance, a ``best'' logistic regression forecasting model is selected. 

Beneath these operational steps, however, there are many details and a substantial statistical foundation that provides a rigorous rationale. We address that rationale in next five subsections with a more technical treatment available in Appendix B. The first subsection addresses the data generation mechanism, which is ``assumption lean'' compared to conventional regression, which is ``assumption laden'' (Buja et al., 2014). This background is necessary to understand what a forecast is estimating. The second subsection considers regularization that is needed to reduce the number of columns of $\Phi(\textbf{X})$. The problem for regression applications is that the number of expansion terms can be equal to or larger than the number of observations. Principle components analysis provides a solution. The third subsection introduces the ``kernel trick'' that allows the regularization to proceed even though $\Phi(\textbf{X})$ is not known. As already noted, the trick depends on $\K = \phiX\phiX^\top$, and $\phiX\phiX^\top$ differs fundamentally from $\phiX^\top\phiX$. The fourth discusses how the relative costs of forecasting errors are properly introduced in classification exercises. The fifth examines the role of tuning parameters and how to obtain valid statistical inference along with honest measures of regression performance. 

\subsubsection{The Data Generation Mechanism}

Conventional regression treats all predictors as fixed, and all results are then conditional on the predictor values in the data. Often this is not responsive to how the data were generated or to the empirical questions being asked. An alternative treats the predictors as random variables, and regression procedures are altered accordingly (White, 1980; 1982; Buja et al., 2014; Berk et al., 2014).

We will be using kernel regression methods as forecasting tools. A defining feature of all forecasting is that the predictors are not fixed. New observations for which forecasts are needed materialize, typically on some regular basis. We must proceed, therefore, within a framework that treats all predictors as random variables. Several key points follow. Details can be found in the paper by Buja et al.(2014).

The data on hand are viewed as a collection of random realizations from a joint probability distribution. The joint probability distribution has mathematically defined expectations, variances, and covariances. As a result, one can treat the joint probability distribution as a population. Alternatively, one can think of all possible realizations of the  random variables as the population.

The realized random variables can be (and usually are) a subset of the random variables that constitute the joint probability distribution. There is the prospect of ``omitted variables.'' Moreover, the variables chosen to be predictors (i.e., \textbf{X}) and the variable chosen to be the response (i.e., $Y$), are determined by a researcher's interests and subject-matter expertise. There is nothing about the joint distribution in itself that determines which random variable should be the response variable and which random variables should be the predictors nor what the functional forms connecting the two should be. In short, it is very difficult to make a convincing case that any statistical formulation derived from those variables is specified correctly. It follows that the proper estimation target usually cannot be some ``true'' response surface, but only an approximation of that true response surface, whose fidelity with respect to the ``truth'' is unknown. This presents no special problems for forecasting because the goal is to arrive at the most accurate forecasts possible with the data on hand. We will see shortly that the forecasts can have good statistical properties. 

\subsubsection{Reducing the Number of Expanded Predictor Terms}

We have denoted the realized predictors by $\textbf{X}$ and their linear basis expansions by $\Phi(\textbf{X})$. Suppose that for some $\Phi(\textbf{X})$, $q \ge N$. Indeed, for many kernels, including the ANOVA kernel, the number of terms in the basis expansion can be infinite. When $q \ge N$, $\Phi(\textbf{X})$ is not a viable regressor matrix. The full set of regression coefficients cannot be not uniquely determined. 

There are several popular ways to reduce the number of expansion terms in $\Phi(\textbf{X})$ or to down-weight them accomplishing much the same thing. One option is to apply a conventional principle components analysis (PCA) to $\Phi(\textbf{X})$. The resulting $N$ principle components (PCs) would be linear combinations of the expansion terms in $\Phi(\textbf{X})$ constructed to be uncorrelated with one another and collectively to incorporate all of the predictive information in  $\Phi(\textbf{X})$, or more technically, in its covariance matrix. How this is accomplished is addressed in detail in Appendix B. 
 
PCs can be ordered from high to low by their contribution to the variance of the set of expanded predictors. Typically, only a leading subset of the $N$ principle components is used in a regression analysis because the leading PCs capture most of the expanded predictor variance. As a result, the number of PCs included can be less than $N$. The $q \ge N$ problem is circumvented. When principle components of $\Phi(\textbf{X})$ are used as a regressor matrix,  we call the resulting regression procedure Kernel Principle Components Regression (KPCR) or, for a binary outcome, Kernel Principle Components Logistic Regression (KPCLR). 

Another recent regularization development  is ``penalized regression.'' The basic idea is to include a penalty for model complexity as part of the fitting function being minimized. The more complex the estimated model, the larger the penalty can be. This makes the estimated regression coefficients smaller in absolute value and the fitted values more smooth. An important and somewhat counterintuitive benefit can be fitted values that have better \textit{out-of-sample} performance, which can be especially important in forecasting applications. However, unlike principle components regression, no predictors need be completely dropped from the regression.
 
We have found our KPCLR approach to be more robust then penalized regression for the kinds of problems often endemic in criminal justice data: highly unbalanced outcomes, long tailed predictor distributions, very different costs for false positives compared to false negatives, and a large number of highly correlated predictors. We will employ the KPCLR approach in our application.

\subsubsection{The Kernel Trick}

The idea of applying PCA to $\Phi(\textbf{X})$ might seem like a routine instance of conventional multivariate statistics. However, $\Phi(\textbf{X})$ must be known. In empirical settings, it rarely is. One then has no way to determine the relationship between $q$ and $N$ and no way to apply PCA to $\Phi(\textbf{X})$.\footnote
{
Recall that the use of linear basis expansions means that $q \ge p$, usually substantially larger, and $q$ can easily be larger than $N$. There is no way to know for sure, but prudence dictates that one assume the worst. Some kind of dimension reduction procedure should be applied.
} 
We appear to be at a dead end. But let's look a little deeper. 

Consider again Equation~\ref{eq:basis} for which there can now be more than one predictor. Because the data on hand are composed of random variables, the estimation target is a conditional expectation of the response, not a conditional mean. This can affect assessments of uncertainty. Formally,
\begin{equation}
\cexpe{Y}{\textbf{\textbf{X}}}= \sum_{m=1}^{q} \beta_{m}\phi_{m}(\textbf{X}),
\label{eq:target}
\end{equation}
where $Y$ can be quantitative or binary, the columns of $\textbf{X}$ contain $1,2, \dots, p$ predictors, and there are $q$ linear basis expansion terms constructed from the full set of original predictors. In principle, one can obtain estimates of $\beta_{0}$ and each $\beta_{m}$ that have the usual desirable properties. The same holds for $\hat{Y}$. Were $Y$ quantitative, one might apply ordinary least squares. Were $Y$ binary, one might maximize the likelihood function. 

However, one must specify each of the basis expansion terms for each predictor. This is a daunting task unless there was credible subject-matter theory specifying the expansion for each predictor and data available to implement those expansions. To take what may seem like a simple example, what is the correct set of expansion terms for the relationship between the age of an offender and recidivism? The relationship is well known to be negative and nonlinear (Berk, 2012), but there is a very large number of potential nonlinear functions. Moreover, the particular expansion terms used will depend on the kind of recidivism (e.g., all new arrests versus all new felony arrests) and features of the offender (e.g., gender). 

Kernel methods respond in a remarkable way. As we address in substantial technical detail in the appendix, there is a formal mapping from the original predictors to a set of linear basis expansions to a particular kernel. The first step from the predictors to the basis expansions was initially discussed in connection with Figure~\ref{fig:d}$: \X \rightarrow \Phi(\X)$. But when kernels were introduced in Section 2.2, the mapping went directly from $\X$ to $\K: \X \rightarrow \textbf{K}.$ What happened to the intervening $\Phi(\textbf{X})$?

The kernel trick (Hastie et al., 2009: 660), justifies proceeding directly from the original predictors contained in $\textbf{X}$ to $\textbf{K}$ because $\K = \phiX\phiX^\top$. Thus, there is no need to ever compute the linear basis expansions. The information they contain is incorporated in $\textbf{K}$ and is sufficient for regression.  

It follows that within the kernel framework, the basis functions themselves are unknown and unrecoverable. They are locked up in the black box. But then, how can one determine if the linear basis expansions are any good? 

Kernels are designed to incorporate \textit{a priori} a very rich menu of expansion terms, often hand-tailored for particular applications. Popular kernels typically are battle-tested in real scientific and policy settings. The ANOVA kernel we favor for regression is one example. Moreover, if the goal is forecasting, a kernel is judged by its forecasting accuracy. Is the accuracy good enough to usefully inform the decisions to be made and more accurate than competing forecasting procedures that rely on functional forms specified using subject-matter knowledge?  We consider these issues in the application presented later.

One might still worry that without $\Phi(\textbf{X})$, important information is lost. If the primary goal is to understand better \textit{why} the predictors are related to the response, the loss is important. Explanation is severely compromised. If the primary goal is forecasting accuracy, the loss may well be irrelevant. The kernel matrix $\textbf{K}$ incorporates the \textit{predictive} information contained in $\Phi(\textbf{X})$. 

At the same time, the values in the kernel matrix can be instructive. Consider the first column of an ANOVA kernel as an illustration. The values in that column are the set of similarities the first observation has with all other observations. To what degree do other observations have a profile like the first observation? Then one can use regression to determine if parolees who are more similar to the first parolee more likely to be arrested? In effect, one shared profile potentially associated with parole failure is being identified. For the second column, there is similar reasoning: are parolees who are more similar to the second parolee more likely to be arrested? All other columns can be interpreted in the same fashion. The kernel trick can help provide answers to such questions, but does \textit{not} reveal what those profiles actually are. This is a consequence of the black box.\footnote
{
There is interesting work that attempts to open the black box (e.g., Grosse et al., 2012; Duvenaud et al., 2013) so that the ways in which linear basis expansions are captured in \textbf{K} are better understood. But we have not seen these insights applied to the ANOVA kernel. }

To summarize, one would like to reduce the number of columns in $\Phi(\textbf{X})$. Principle components analysis is one good option. However, $\Phi(\textbf{X})$ is not available. One only has $\textbf{K}$. Thanks to the kernel trick, one can apply principle components analysis to $\textbf{K}$ to arrive at the desired result. 

\subsubsection{Introducing the Relative Costs of False Positives and False Negatives}

In criminal justice policy settings, the costs of false negatives and false positives will generally be different. For example, when a release decision needs to be made at arraignment, there are two kinds of mistakes that can be made. An individual is released and then fails to appear at a subsequent court hearing or an individual is not released and would have appeared. The consequences and costs of these mistakes are rather different and should be built into the forecasts of failure to appear; they should affect the forecasts themselves (Berk, 2012). 

Forecasting procedures can differ dramatically in the mechanisms by which the different relative costs of forecasting errors are be introduced. For conventional logistic regression, the key is a threshold on the fitted values; values larger than some threshold imply that the associated observations belong in one outcome class, and fitted values equal to or smaller than that threshold imply that the associated observations belong in the other outcome class. A widespread ``default'' is to ignore the possibility of asymmetric costs and employ a threshold of .50. 

If the logistic regression model meets all of the conventional assumptions, the fitted values can be interpreted as asymptotically unbiased probability estimates. Then, if estimated probabilities are greater than .50 one outcome class is forecasted, and if estimated probabilities are equal to or smaller than .50 the other class is forecasted. Implicit is that the costs of false positives and false negatives are exactly the same. 

A variety of relative costs can be easily introduced using different thresholds. Suppose there are two outcome classes following an arraignment, arrested for a felony (coded ``1'') or not arrested for a felony (coded ``0''). One might call the arrested class a positive and the non-arrested class a negative. Here, the 1/0 values and the terms ``positive'' and ``negative'' are assigned arbitrarily. A false positive would be incorrectly forecasting an arrest. A false negative would be incorrectly forecasting the absence of an arrest. Then, suppose stakeholders determine that false negatives are twice as costly as false positives. Imposing a threshold of .33 on the fitted values forces the false negative to false positive cost ratio of 2 to 1 on the forecasted class (i.e., $.67/.33$). That is, for a case to be forecasted as an arrest, its fitted value must be in excess of only .33. If not, the forecasted class is the absence of an arrest. It is ``easier'' to forecast an arrest, which is consistent with the stated cost ratio. In contrast, a threshold of .75 implies a cost ratio of 1 to 3 (i.e., $.25/.75$). False positives are now three times more costly. It is ``harder'' to forecast an arrest. 

For KPCLR, the introduction of relative costs is done somewhat differently. When a logistic regression is run, the data are weighted so that the marginal distribution of the response is altered, and the logistic regression is fit with these weights. Suppose a positive is now an FTA, a negative is now the absence of an FTA, and false positives are taken to be twice as costly as false negatives. Case weights are given to the logistic regression so that \textit{actual} positives have twice the weight of \textit{actual} negatives.\footnote
{
The weights are introduced as a vector of length $N$ with values that leave the effective sample size unchanged. (i.e., They have a mean of 1.0.)
} 
This does not immediately produce results in which false positives are twice as costly as false negatives. As we describe in more detail shortly, the KPCLR fitting algorithm we employ gradually increases the complexity of the fitted values. When the increasingly complex fitted values arrive at the specified cost ratio, one can have the requisite cost-ratio result. 

\subsubsection{Tuning and Statistical Inference}

Researchers are gradually coming to realize that model selection is not without its inferential perils. Conventional approaches in which model selection and statistical inference are undertaken with the same data risk serious bias in parameter estimates and highly misleading confidence intervals and statistical tests (Leeb and P\"{o}tscher, 2005; 2006; Berk et al., 2010). Unfortunately, the importance of tuning parameters puts us squarely in the middle of these problems. We have found no practical solutions besides using split samples.\footnote
{
The recent literature is quite rich, but key problems are not yet solved (Berk et al., 2013; Lockhard et al., 2013; Voorman et al. 2014).
}
Split sample approaches properly implemented promise valid statistical inference at the price of reduced statistical precision (Berk et al., 2010) and more complicated data management. Often, this is a tradeoff well worth making (Faraway, 2014). The basic idea is to use different random subsets of the data for different data analysis tasks. Model selection, parameter estimation, and model forecasting performance are not undertaken with the same data. 

Our approach has some novel features. We will proceed sequentially using the following operations.
\begin{enumerate}
\item
The data are randomly split into three disjoint subsets we will call \textit{training data}, \textit{validation data} and \textit{test data}.
\item
Stakeholders supply relative costs of false positives and false negatives, and these two costs are used to construct case weights for the \textit{training} data. 
\item
A set of promising ANOVA kernels is specified with different tuning parameters ($\gamma$ and $d$) from which KPCLR models will be built.\footnote
{
For reasons discussed earlier, we favor ANOVA kernels for regression applications.
} 
\item
For each kernel, a set of proportions of variance explained is defined (i.e., values for $\rho$) that will be used to determine the number of principle components provisionally included (e.g.,$\rho=35\%, \rho=40\%, \ldots, \rho=95\%$).
\item
For each kernel in step 3 and each value of $\rho$ in step 4, a KPCLR model is built with the \textit{training data}. 
\item
Using performance with the \textit{validation} data as a guide, one preferred KPCLR model is chosen applying procedures to be explained shortly.
\item
Forecasting accuracy is determined for the model selected in Step 5 by predicting into  \textit{test} data using the preferred regression model built from the \textit{training} data.
\end{enumerate}

These steps come bundled with many demanding particulars. To begin, there is currently no clear statistical guidance on the relative sizes of split samples, even when there are only two (Faraway, 2014). A lot depends on features of the data and the models being used. Samples of equal size are often reasonable.

The weights used in the logistic regressions are determined \textit{a priori} by the cost ratio of false positives and false negatives. Once a cost ratio is determined, the weights follow as a simple mathematical exercise. For our analysis of FTAs, false positives are taken to be twice the cost of false negatives. Therefore, all actual non-FTA cases are given twice the weight of all actual FTA cases, scaled so that the effective sample size is unchanged. 

Sets of values for the ANOVA kernel parameters $\gamma$ and $d$ must be specified. This will be easier if all predictors in $\textbf{X}$ are standardized as z-scores.\footnote
{
If the predictors are not standardized, the values of the turning parameters can be dramatically affected by the units of measurement, which are here a distraction.
}
For example, $d$ might be 2, or 3, and initial values for $\gamma$ could range from .01 to 100 as multiples of 10: .01, .10, 1.0, 10, 100. The search could then become more concentrated around the most promising initial values. Because the models are evaluated with out-sample-performance, the primary penalty of model searching is computer time. We have found that a consideration of twenty to thirty models can be sufficient. We will have an example shortly. 

The number of principle components needed as regressors must be determined empirically by fitting models with different proportions of variance explained (e.g. $\rho = 30\%, 35\%, \ldots, 95\%$). For each combination of $d$, $\gamma$, and $\rho$, a KPCR model is built. This may seem quite daunting, but with our software, a rich set of models can be produced in a matter of minutes, not hours. 

Model evaluation follows. For each model, we provide (1) the number of false negative errors and the number of false positive errors in the \textit{validation} data, (2) the aggregate cost-weighted error for the predictions on the \textit{validation} data and (3) the proportion of PCs used. A first cut eliminates all models whose cost ratio of false negatives to false positives is not sufficiently close to the \textit{a priori} cost ratio. Such models are not responsive to stakeholder policy preferences. A second cut eliminates all models with an adequate cost ratio, but disappointing forecasting accuracy. Among the models that make the second cut, preference is given to models that use fewer principle components. There seems to be no point in wasting degrees of freedom. 

There are some tricky issues at this stage because of the temptation to treat each KPCLR as a usual linear model popular in the social sciences. None are. Once the regressors are kernelized, logistic regression becomes a black box algorithm from which to construct useful forecasting procedures. There is no intent nor capability to reveal the subject-matter mechanisms by which the response is related to the original predictors. In addition, most of the usual regression diagnostics are not relevant and can even be misleading. One key reason is that we are interested in forecasting accuracy for each of the outcome classes (e.g., fail or not fail), not the logistic regression fitted values. Yet, most regression diagnostics build on the in-sample fitted values, often treated as probabilities. 

The \textit{test} data, the third random split, is then used to provide for the preferred model an honest assessment of forecasting accuracy not subject to overfitting. These are out-of-sample assessments because the third split had no role when constructing the model and no role when selecting the best model. Nevertheless, there may be residual concerns about overfitting when earlier the ``best'' KPCLR model is selected. Just as in boosting of binary outcomes, the iterative process can produce dramatic overfitting. Yet in general, overfitting of the fitted value in \textit{training data} can actually benefit \textit{out-of-sample} forecasting accuracy (Mease et al., 2008). 

This remarkable result makes sense in the machine learning world of classification. With a rich menu of predictor transformations summarized by an increasing number of principle components, any training data fit will improve until it can improve no more. That fit will have two components. The first is systematic features of the training data that may normally be very hard to find, but can be captured by a sufficiently rich kernel expansion and a sufficient number of its principle components. The second is idiosyncratic patterns in the training data swept up in fitting process that are not features of the joint distribution from which the data came. The latter are often characterized as \textit{chance variation}. When people speak of the dangers of overfitting, they are referring to the chance variation only. 

If one constructs a forecasting procedure and evaluates its performance using the same data, the two components cannot be disentangled. There is, then, a genuine reason for concern. But if there are data not used to build the forecasting procedure that can be used for performance assessments, one can obtain honest performance evaluations without the contaminating effects of overfitting. One is then left with the benefits of the hard-to-find, systematic features of the data. 

We capitalize on just such thinking. KPCLR models are undertaken with the \textit{training} data, in which overfitting can be a virtue. KPCLR model selection is done with the second split, the \textit{validation} data to help counteract the misleading properties of in-sample assessments. Once a KPCLR model is chosen, an honest evaluation of forecasting accuracy is provided by the third split, the \textit{test} data. For the KPCLR model chosen, this evaluation is not a product of overfitting. 

With this split sample approach, one has forecasting tools with good statistical properties. Perhaps most important, one has a forecasting procedure that provides asymptotically unbiased forecasts derived from the population response surface approximation (Buja et al., 2014). But, the price should now be clear. Each subsample will have many fewer observations than the original data set, and forecasting accuracy can be substantially reduced. In addition, any estimator properties that depend on conventional asymptotics can be put in harm's way. In the example we turn to shortly, sample sizes will be relatively large to minimize the consequences of these difficulties. But more generally, proper statistical inference after model selection is a very active research area, and there does not appear to be yet any definitive answers absent significant tradeoffs (Berk et al., 2013; Lockhard et al. 2014; Voorman et al., 2014).

\section{An Application}

We address in this application forecasts of FTAs: a failure to appear in court after a preliminary arraignment in which formal charges are not dismissed. In the jurisdiction from which the data were collected, about 40\% of those who should appear in court at subsequent dates fail to appear. Many criminal justice officials and court observers believe that forecasts of individuals who are likely to miss their court appearances could usefully inform judges' bail and release decisions. We will conduct an illustrative forecasting competition between stepwise logistic regression and KPCLR to see which approach is more instructive. Both procedures will be seriously challenged because as we explain shortly, the forecasting task is difficult.  

\subsection{The Population}

The population for which our forecasting procedure will be developed is individuals from a particular, large metropolitan area released at arraignment with the requirement that they return to court at a later date. The data include all arraignment decisions leading to release awaiting trial from 1/1/2007 through 11/30/2010.\footnote
{
Without a release, it would be impossible to learn how the individual performed ``on the street.''
}
Each element in the population is a single bail decision for a single offender. Individual offenders, court docket numbers, and even criminal incidents can therefore appear more than once. The observational unit is the case rather than the individual because each case requires a release decision.\footnote
{
The facts and outcomes associated with a given case can vary dramatically even for the same offender. Moreover, repeat cases are somewhat rare in this instance because of the relatively short time period that defines the population. Finally, evaluating forecasting performance out-of-sample, further helps to mitigate any estimating problems that could result. }

From existing electronic criminal justice records, we have the usual background variables such as age, gender, and prior record. We have the types of charges heard at the arraignment. And, we have each defendant's prior record as a juvenile. For prior record as either an adult or a juvenile, we have the date on which each arrest occurred. Finally, we have for two years after the release all arrests and any failures to appear in court. 

Although an FTA is a violation, it often has a different etiology from conventional street crimes. Bornstein and his colleagues (2013) argue that FTAs often result from a faulty memory, an inability to get timely transportation, household responsibilities such as child care, or work-related obligations. Views about the fairness of the adjudication process can matter as well. They also show that written reminders can reduce FTA rates, especially if the reminder provides information about the negative consequences of a failure to appear. 

Unfortunately, among our potential predictors we have virtually no measures of such factors. It follows that our ability to accurately forecast FTAs is seriously compromised from the start. We stress, however, that this application was not selected to make any particular methodological point. We had no idea how good our forecasts would be before they were made. Nor did we know in advance how the competing forecasting methods would perform. The forecasting assignment and the data were brought to us by real stakeholders who were seeking technical assistance. In retrospect, however, the forecasting exercise is ideal for our purposes. When a forecasting task is easy, most any forecasting procedures will do well. Difficult forecasting tasks allow one document the relative performance of different methods. 

Some cases in the data had to be discarded. In a few instances, there were serious problems with missing information. There were also cases in which an individual was incarcerated in a local prison post arraignment. There were, as well, instances in the two years after release in which an individual spent 18 months or more in jail because of detainers or holds to face other charges. Under both circumstances, there was virtually no opportunity to succeed or fail during the two year follow up. Such cases were dropped. In the end, there were 175,361 observations in the data set.

We proceed as if there is a joint probability distribution from which the 175,361 cases were realized or alternatively that the data are a random subset of all possible cases that could be realized. This perspective is essential because forecasts of future cases only makes sense if they are realized in the same manner. In this application, the joint probability distribution is just a statistical summary of the social processes that send cases to arraignment in this metropolitan area for the years around the time the data were collected. 

\subsection{The Working Sample}

One can consider the 175,361 cases for which there are both measured predictors and outcomes as a training sample. But our intent is to illustrate how stepwise logistic regression and KPCLR can perform even on samples of modest size, when there is no a priori model and when many key predictors are unavailable. Consequently, we will work with a random sample of 1500 cases drawn from the 175,361. Because we will be using a split sample approach, we randomly subset the 1500 cases into three random samples of 500 for the procedural steps described in Section 2.5. For these three samples, it may be conceptually easier to consider the set of 175,361 cases as the (finite) population.

\subsection{Variables}

There are 41 predictors. We have a few biographical variables such as age, and gender, but most come from adult and juvenile rap sheets and current charges.\footnote
{
The rap sheet data only include arrests from within the state in which the metropolitan area is located. But most crime, like most politics, is local. Relatively few arrests are missed. Moreover, the impact of priors on forecasting accuracy is highly nonlinear. Although there are special concerns about ``frequent flyers,'' variation of several prior arrests (e.g., 35 arrests versus 40 arrests) is for such offenders unrelated to forecasting accuracy. If out-of-state arrests matter for prediction, it is for offenders with very few in-state arrests. But then, there is some evidence that other predictors pick of the slack.
} 
Many are correlated with one another, often strongly. Logistic regression can be sensitive to high multicollinearity, but with our main interest in forecasting accuracy, it does not cause serious problems. Fitted values are less affected by multicollinearity than estimates of the regression coefficients, and the dimension reduction tools we apply moderate much of the remaining instability.

As already noted, there is virtually no information on routine life circumstances that might influence a failure to appear. This makes the forecasting challenge substantial. Ideally, the variables we are able to include can serve at least in part as proxies for more important omitted predictors. The full set of predictors is listed in Appendix A.

\subsection{Cost Ratios}

Forecasting binary outcomes will almost inevitably result in some false positives and some false negatives. For this forecasting application, a false positive is incorrectly predicting that an individual will fail to appear in court when ordered to do so. A false negative is incorrectly predicting that an individual will not appear in court when the individual actually would. For criminal justice stakeholders, both forecasting errors are undesirable and both have costs. When researchers accept the default procedures provided by the usual logistic regression software, they are accepting a default fitting function is which the two costs are treated the same: the costs are symmetric, and their cost ratio is 1 to 1. This is usually inconsistent with the preferences of stakeholders who will use and be affected by the forecasts.

At the time these data were collected, the metropolitan area were operating under serious resource constraints. There was insufficient jail space should the forecasts produce a large number projected FTAs cases requiring incarceration. A key implication was that the false positives could be in the aggregate very costly. At the very least, there could be serious ``overcrowding.'' Less costly options were being evaluated, but many had little demonstrable impact or were burdened by legal and political constraints. For example, methods that have been used to monitor individuals on probation may be turn out to be illegal for defendants who have yet to be adjudicated. Provisionally, therefore, false positives are assumed to be more costly than false negatives.  For the purposes of illustration, we will use a cost ratio of false positives to false negatives of 2 to 1 that is broadly consistent with stakeholder preferences. 

\subsection{Results for Stepwise Logistic Regression}

All of the available predictors were used in a stepwise selection (backward elimination using the AIC) applied to the \textit{training} split. The resulting smaller model with 17 predictors was applied to the second random split of the data to obtain parameter estimates.\footnote
{
We used the stepwise regression in \texttt{R} \textit(stepAIC), which is part of the \texttt{MASS} library. 
}
Using the validation split to obtain parameter estimates eliminates concerns about model overfitting and the distortions of statistical inference that can follow from model selection. Assessments of forecasting accuracy were undertaken using the (third) test split and the model obtained from the second split. 

The fitted values from the stepwise regression were distributed much like the fitted values when all of the regressors were included, just as one would expect. There are perhaps other sensible ways to employ stepwise regression with split samples, but the approach we used avoids the usual errors associated with single sample approaches and is consistent with spirit of typical criminal justice applications. 

\begin{figure}[htp]
\begin{centering}
\includegraphics[scale=.30]{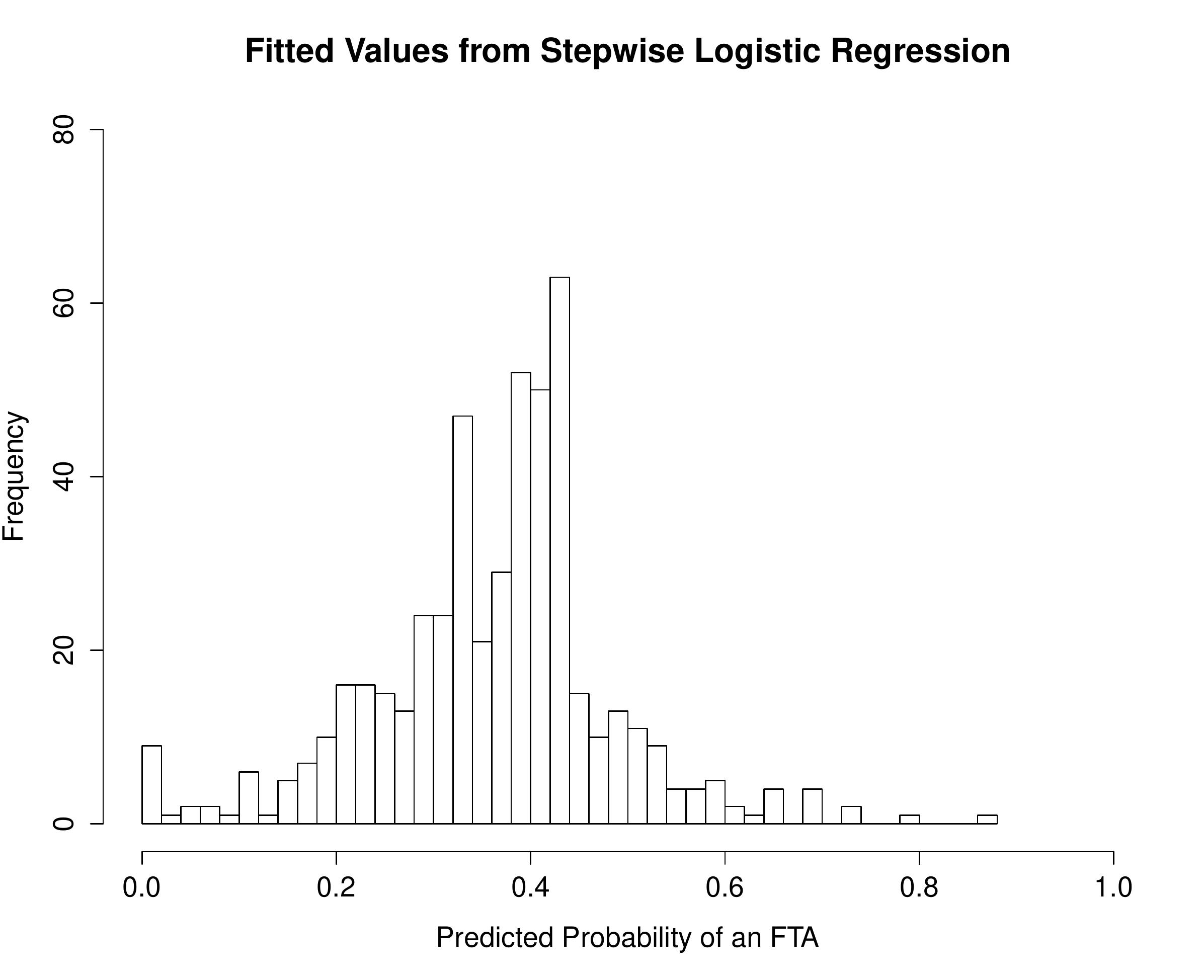}
\caption{Histogram for Out-of-Sample Stepwise Logistic Regression Fitted Values (N=500)}
\label{fig:s}
\end{centering}
\end{figure}

Figure~\ref{fig:s} shows the fitted values from the \textit{test} data centered a little below .40. The median is .38, the first quartile is .30, and the third quartile is .42. But there are also a few cases near 0.0 and .90. To arrive at forecasted classes, a threshold on those fitted values is required. Using the 2 to 1 cost ratio, that threshold is .67. ($.67/.33 \cong 2$). Note, it can be misleading to interpret the fitted values as conventional probabilities. Given the likely omitted variables and functional forms that essentially determined by convenience, the regression model is badly misspecified. The conventional assumptions of logistic regression are not met. 

Table~\ref{tab:L} is the resulting ``confusion table,'' a cross-tabulation of the actual outcome class by the forecasting outcome class. Stepwise logistic regression is able to correctly identify in advance 99\% of those who appeared in court when they should. But, less than 3\% of those who failed to appear could be correctly identified in advance. Of the 500 cases, including 198 actual FTAs, only 6 were forecasted as an FTA. The forecasts amount to always forecasting that an individual will appear in court when ordered to do so. The predictors are of almost no help.

One might argue that the model specification is too simple. Even within the limitations of the available predictors, important interaction effects and non-linearities have not been included. Had they been, forecasting performance would have been better. This argument has merit were there credible subject-matter theory and earlier empirical research providing guidance for how the model specification could be improved \textit{with the predictors on hand}. We have not found such guidance. 

\begin{table}[htp]
\begin{center}
\begin{tabular}{| c | c | c | c | }
\hline \hline
~ & Predict No FTA & Predict FTA & Model Error \\ \hline
Actual No FTA & 300 & 2 & 0.01 \\ 
Actual FTA & 192 & 6 & 0.97 \\ \hline
\hline
\end{tabular}
\caption{Failure to Appear (FTA) Stepwise Logistic Regression Confusion Table Constructed from the \textit{Test} Data (N = 500)}
\label{tab:L}
\end{center}
\end{table}

\subsection{Results for KPCLR}

We proceeded with the very same three random splits of the data applying the procedures of Section 2.5, Steps 1-6, and searching over a parameter grid with $d = 2$ or $d=3$ and $\gamma$ values of  .01, 1, 3, 1. Four candidate ANOVA kernels were seriously considered: $(\gamma, d) = \{(0.1, 2),~ (3, 2), ~(0.1, 3), ~(3, 3)\}$. For each of the candidate kernels, the values of $\rho$ were fixed at $30\%, 35\% \ldots, 95\%$. Each kernel's performance in a logistic regression was judged as principle components were added in 5\% increments of the kernel variance accounted for. Regression parameters were estimated with the training data, and performance was evaluated using the validation data. Forecasting performance was assessed with the test data.\footnote
{
We also briefly tried $\gamma$ values of 100 and 300 with $d$ equal to 2 or 3, mostly out of curiosity. These kernels performed about the same as the better models ultimately selected but were discarded because they led to more localized support than was probably necessary.  
}

\begin{figure}[htp]
\begin{centering}
\includegraphics[width=5.0in]{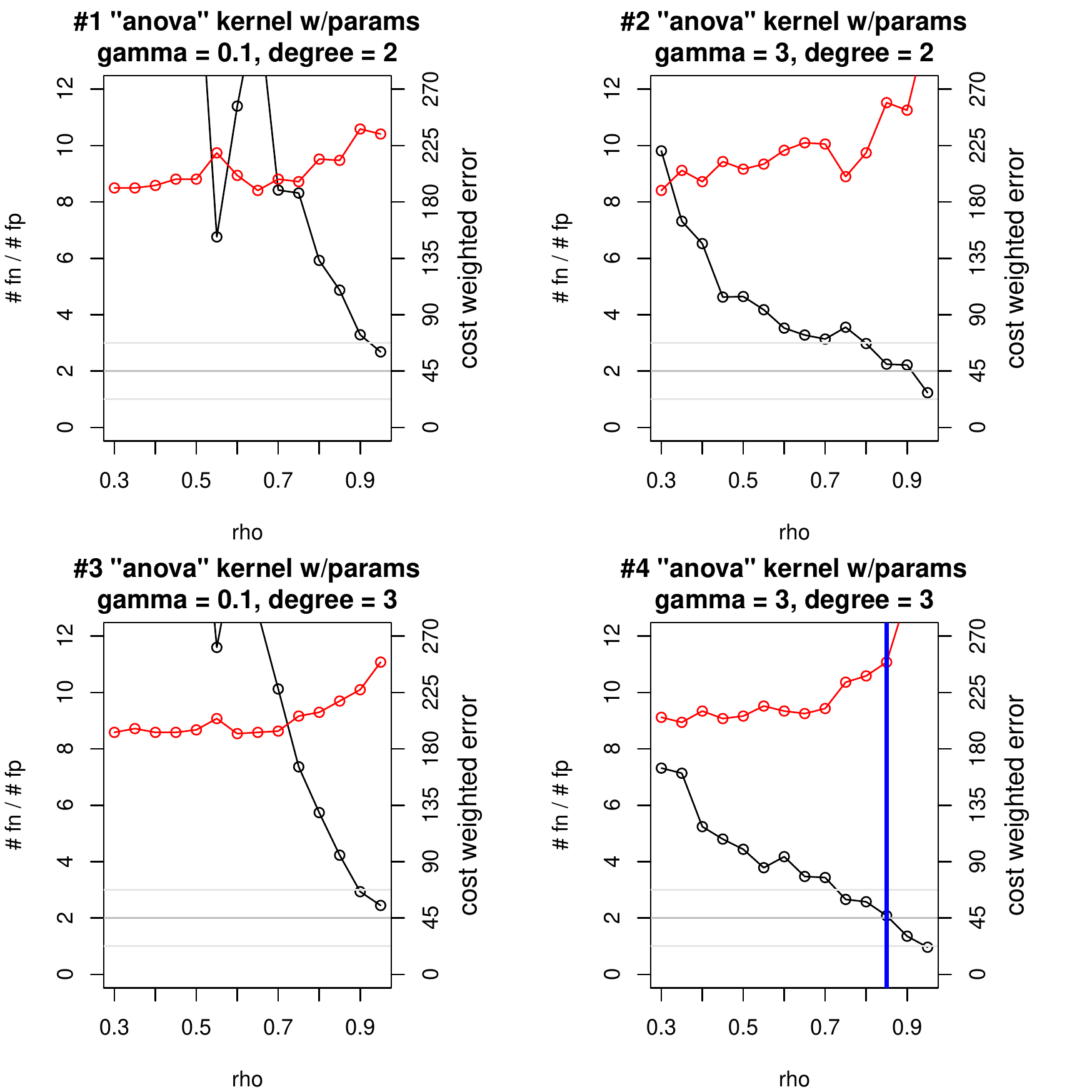}
\caption{Model Performance as a Function of  $\rho$}
\label{fig:t}
\end{centering}
\end{figure}

Figure~\ref{fig:t} shows the diagnostic plots used to determine which model is preferred (Step 6 in Section 2.5). The horizontal axis represents values of $\rho$, the fraction of variance of the transformed predictors accounted for by the principle components. The larger the value of $\rho$, the more PCs were used as predictors. The left vertical axis and black line show the ratio of the number of false negatives to the number of false positives. Because the target cost ratio of false positives to false negatives is 2 to 1, the goal was to arrive at empirical results in which there were two false negatives for every false positives because false positives were twice as costly. The right vertical axis and red line show the cost-weighted number of forecasting errors in the validation data.  Finally, the vertical blue line in the graph on the lower right, shows the value of $\rho$ for the selected kernel.

The two kernels on the left fail by the first criterion. No matter the value of $\rho$, the required cost ratio of 2 to 1 is never achieved, and to even get close requires using virtually 100\% of the PCs. The two kernels on the right arrive approximately at  the desired cost ratio. And when that is achieved, cost-weighted forecasting error is about the same (i.e. approximately 240). We chose the kernel at the lower right because it achieves comparable performance using fewer PCs, but in practice, either kernel will forecast with about the same level of skill. Note that in both cases, one can achieve substantially better forecasting accuracy with far fewer PCs, but the cost ratio is far from 2 to 1. Those forecasts are not responsive to stakeholder policy preferences and consequently, these models are not acceptable for use in practice. 

\begin{figure}[h]
\begin{centering}
\includegraphics[width=3in]{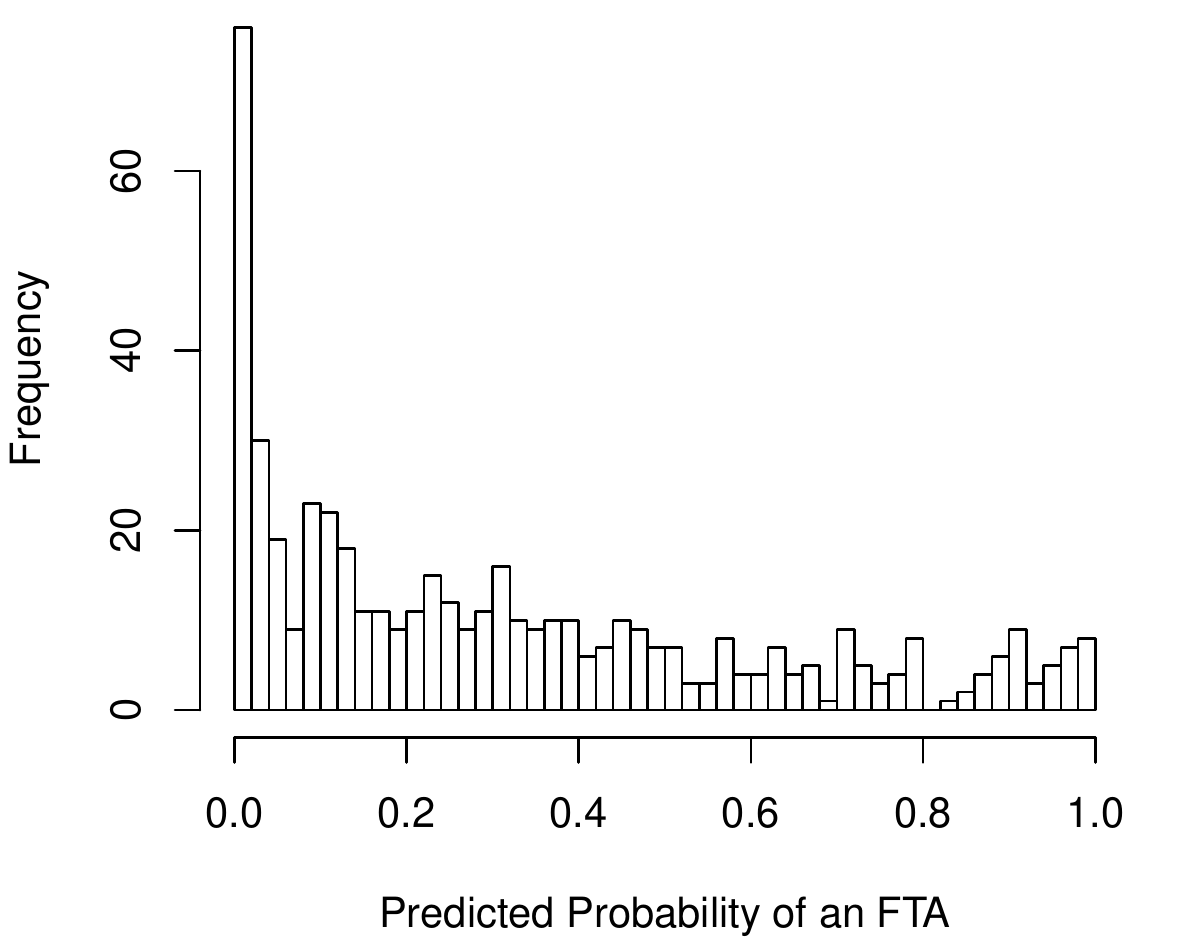}
\caption{Histogram for Out-of-Sample Fitted Values from the Principle Components
Kernal Logistic Regression}
\label{fig:f}
\end{centering}
\end{figure}

Figure~\ref{fig:f} shows the fitted values when the ``best'' regression is applied to the test data. Compared to the fitted values from the stepwise logistic regression, the fitted values are far more dispersed with much thicker tails, especially at the low end. Greater distinctions are being made between defendants; the fitted values discriminate better. Moreover, because the fitted values are asymptotically unbiased estimates of the fitted values in the population response surface approximation, they can be treated as probabilities. 

Table~\ref{tab:k} shows the out-of-sample forecasting results from the test data. 81\% of the non-FTAs are correctly identified in advance. 31\% of the FTAs are identified in advance. This is a dramatic improvement over the forecasts from the stepwise regression. The cost ratio of false positives to false negatives approximates 2 to 1 (i.e., 141/57= 2.47 which is within 25\% of the target). In general, it is virtually impossible to hit the cost ratio exactly because, as required, the data were not tuned using the test data. But because we nearly had an exact hit  with the validation data, there is not cause for concern. 

\begin{table}[h]
\begin{center}
\begin{tabular}{| c | c | c | c | }
\hline \hline
~ & Predict No FTA & Predict FTA & Model Error \\ \hline
Actual No FTA & 239 & 57 & 0.19 \\ 
Actual FTA & 141 & 67 & 0.69 \\ \hline
\hline
\end{tabular}
\caption{Failure to Appear (FTA) Confusion Table Constructed from \textit{Test} Data (N=500) for the KPCLR Procedure.}
\label{tab:k}
\end{center}
\end{table}

In these data, nearly 40\% of all arraigned individuals who are not held in jail actually fail by an FTA. Table~\ref{tab:k} indicates that were the courts to release individuals forecasted to return to court when ordered to do so, 37\% would fail by an FTA (141/(141+239) = .37). Clearly, this is a small improvement. But just as clearly, it is a policy-relevant improvement for a court system that arraigns over 50,000 individuals a year. If our results we used to determine whom to release, there could be approximately 1500 fewer FTA incidents. And the point is more general. Small relative improvements in the performance of large criminal justice systems can translate into dramatic absolute gains that can make an important difference.\footnote
{
We have not considered here the possibility of additional benefits of requiring at least some individuals to post a bond before release. That is, these figures necessarily represent the policy status quo.
} 

It is also possible to do much better. It cannot be overemphasized that the 37\% failure rate among those predicted to succeed results substantially from the 2 to 1 cost ratio. If as a policy matter, false positives were given less weight, more true negatives would be correctly identified, and the projected FTA failure rate would be reduced, perhaps dramatically.  

According to stakeholders, the 2 to 1 cost ratio of false positives to false negatives is at this point highly provisional, and small changes in the cost ratio can make a big difference in forecasting accuracy. For example, some alternative analyses suggest that a cost ratio of 1 to 1 could reduce the failure rate among those forecasted to succeed from 37\% to nearly 30\%. The number of FTAs is now reduced by about 5000. Other improvement would follow from using larger samples and better measures the life circumstances associated with FTAs. 

There is also an important statistical message. The results from the stepwise logistic regression and KPCLR were dramatically different even though they were applied to the same data, analyzed with the same split-sample approach, had access to the exact same predictors, and were subject to the same cost ratio. Comparative assessments were fair because they were undertaken on the same out-of-sample \textit{test} data. Yet, important predictive differences between the FTA cases and the non-FTA cases were discovered and exploited by the kernel approach that were missed by the stepwise logistic regression. The experience comparing logistic regression to machine learning is similar (Berk and Bleich, 2013).

\section{Conclusions}

Even with a very weak set of predictors, FTAs and non-FTAs can be forecasted with better accuracy than produced by current practice in at least one large metropolitan area. One could do substantially better with cost ratios that increased the relative costs of false negatives. With information about defendants' life circumstances and views of the criminal justice system, still more accurate forecasts are likely. Finally, there is the possibility of interventions, such as mailed or texted reminders, that could reduce the number FTAs and make them more predictable. 

As a technical matter, kernelized principle component logistic regression seems to be a useful alternative to conventional logistic regression when a researcher's primary interest is in classification and forecasting. By constructing in advance a rich menu of basis functions, complicated non-linear relationships and interaction effects can often be captured. Kernelized regression attempts to anticipate complexities that inductive methods like random forests discover on-the-fly (Brieman, 2001).

A congenial division of labor between the two black box procedures may follow. Kernel approaches seem well suited for ``small data.'' Machine learning approaches seem well suited for ``big data.'' The $N \times N$ dimensions of the kernel matrix make KPCR and KPCLR computationally impracticable for large data sets, and inductive machine learning procedures often need large data sets to work most effectively.\footnote
{
For example, the trees grown by random forests cannot be large when the sample is small  because each tree rapidly runs out of observations. 
}
However, the distinction between ``small'' and ``large'' will be situation specific. A lot depends on the number of parameters to be estimated, the number and nature of tuning parameters, and the complexity of the decision boundary. Moreover, the relative merits of machine learning and kernel regression methods have not yet been definitively determined although there is evidence that some variants of kernel regression may be preferred in certain situations (Zhu and Hastie, 2005). There are also interesting proposals to make kernel logistic regression more feasible with large data sets (Qin et al., 2011).

Can conventional logistic regression compete? When the predictor distinctions between outcome classes are uncomplicated, even very simple logistic regression models can perform well. Machine learning and kernel methods may then confer no special advantage. But as others have emphasized (Berk and Bleich, 2013, Ridgeway, 2013, Bushway 2013), one cannot know in advance how complicated the key relationships are. Prudence will often dictate assuming the worst. 

A happier reason to favor conventional logistic regression is when there is subject-matter knowledge indicating in a convincing manner how to properly specify the regression model and data available to actually implement that specification. Then, one can have it all: excellent forecasts and genuine explanatory insight as well. 

This scenario seems unrealistic for the processes by which some people return to court as ordered and some do not. There is probably not a jurisdiction in the United States with the requisite subject-matter knowledge and data. Prospects for both in the medium term are not encouraging. The use of Burgess-like scales constructed from risk factors determined by logistic regression have done yeoman service in the past. It is time to move on.

\section*{Acknowledgements}

We would like to thank Larry Brown, Andreas Buja, Ed George, and Dean Foster for helpful discussions. Adam Kapelner acknowledges the National Science Foundation Graduate Research Fellowship for support.

\section*{Appendix A}
Candidate predictors include the following variables. 
\begin{enumerate}
\item
The age of the offender
\item
The gender of the offender
\item
The zipcode in which the offender lives if it is a high crime zip code
\item
The length of the follow up period
\item
The total number of prior charges as a juvenile
\item
The number of ``serious'' prior charges as a juvenile
\item
The number of ``violent'' prior charges as a juvenile
\item
The number of sex crime prior charges as a juvenile
\item
The number of firearm prior charges as a juvenile
\item
The number of weapon prior charges as a juvenile
\item
The number of drug prior charges as a juvenile
\item
The number of property crime prior charges as a juvenile
\item
Whether there was any prior charges as a juvenile
\item
Whether there was any violent prior charges as a juvenile
\item
The age of the first adult charge while a juvenile
\item
The number of prior murder charges as an adult
\item
The number of ``serious'' prior charges as an adult
\item
The number of ``violent'' prior charges as an adult
\item
The number of sex crime prior charges as an adult
\item
The number of firearm prior charges as an adult
\item
The number of weapon prior charges as an adult
\item
The number of drug prior charges as an adult
\item
The number of property crime prior charges as an adult
\item
Whether there were any charges at the arraignment
\item
The number of murder counts at the arraignment
\item
The number of weapons counts at the arraignment
\item
The number of property crime counts at the arraignment
\item
The number of drug distribution counts at the arraignment
\item
The number of domestic violence counts at the arraignment
\item
The number of violent crime counts at the arraignment
\item
The number of serious crime counts at the arraignment
\item
The number of sex crime counts at the arraignment
\item
The number of firearm crime counts at the arraignment
\item
The number of drug possession crime counts at the arraignment
\item
The number of sex crime counts at the arraignment
\item
The number of prior FTAs
\item
Whether the individual is currently on probation
\item
The number of prior abscondings
\item
The number of prior probation violations
\item
The number of prior days in jail
\item
The number of prior confinement days
\end{enumerate}

\section*{Appendix B: Kernel Principle Components Regression}

\subsection*{Principle Components and the Kernel Trick}

Using subset of principle components (PCs) as regressors is an old regression story (see Hastie et al., 2009: Section 3.5.1), commonly motivated by unacceptably high multicollinearity among the predictors. An $N \times p$ matrix of predictors is transformed into an $N \times p$ matrix of PCs that are orthogonal by construction and thereby uncorrelated with one another. The complete set of $p$ PCs account for all of the variance in the matrix of predictors. The PCs employed in the regression as predictors are then a subset of the $p$ PCs. Often they are a small subset because a small fraction of the PCs can account for most of the variance in the original predictor matrix.

Using a subset of PCs is a form of \textit{regularization} because the fitted values will be less variable than had a larger number of PCs been used. One hopes to introduce only a small amount of bias into estimates of the fitted values for a large reduction in the fitted values' variances. We build on these ideas for kernel principle components regression (KPCR). There is a lot of detail, but we provide a summary near the end. 

As in Section 2, there are $N$ data vectors $\x_1, \ldots, \x_N \in \mathbb{R}^p$, each with dimension $1 \times p$. For our application in Section 3, these vectors correspond to the characteristics of the individual offenders, such as age and number of prior jail terms. As a simple running example to illustrate the conceptual material, suppose that each offender has three covariates: age, number of prior jail terms, and number of drug prior charges ($p=3$).

Imagine there is a function $\Phi: \mathbb{R}^p \rightarrow \mathbb{R}^q$ that transforms $\x$ into $\phix$, a set of $q$ basis terms with $q > p$. Often, $q$ is larger than $N$, and possibly even be infinite. As we explain shortly, this function $\Phi$ need not be explicitly specified or even known, but we have already used for simple illustrative purposes computing polynomial powers of the predictors. If, for instance, $\Phi$ expanded the original covariate vectors to include all polynomial terms up to cubic terms for each of the $p=3$ predictors, then $q$ would be 9. The vector $\phix$ would contain terms such as $\text{age}, \text{age}^2 $, and $\text{age}^{3}$. Therefore, $\Phi$ has expanded the number of predictors available for a regression problem and introduced the flexibility to fit nonlinearities. 

Recall that the sample covariance matrix of the data matrix $\X$ can be defined as:\footnote
{
This is the maximum likelihood estimate of the covariance matrix, not the usual unbiased estimate using $1/(N-1)$,
}

\bneqn 
\bv C' = \frac{1}{N}(\X - \bar{\X})^\top (\X - \bar{\X}) 
\eneqn
where $\bar{\X}$ is the matrix of sample averages duplicated over columns so that $\X - \bar{\X}$ is \qu{centered} i.e. each column's average is 0. The $p \times p$ matrix $\bv C'$ is commonly the input matrix for principle components analysis (PCA). 

In kernel regression, the linear expanded basis $\phiX$ is used as the predictor matrix. Its sample covariance matrix is: 

\bneqn 
\bv C = \frac{1}{N} (\phiX - \overline{\Phi(\X}))^\top (\phiX - \overline{\Phi(\X})) = \frac{1}{N} \widetilde{\phiX}^\top \widetilde{\phiX}. 
\eneqn
where $\widetilde{\phiX}$ is the matrix of expanded bases and is centered analogously to $\X - \bar{\X}$. It is important to note the $\bv C$ \textit{is not} the kernel matrix. It is the sample covariance matrix across the expanded set of predictors. 

\subsection*{PCA as Eigendecomposition}

In order to carry out PCA, one first computes the eigendecomposition of the matrix portion of $\bv C$ given as

\bneqn 
\label{eq"blsh}
\widetilde{\phiX}^\top \widetilde{\phiX} = \bv V \bv \Lambda \bv V^\top = [\v_1 \ldots \v_q] \threebythreemat{\lambda_1}{}{}{}{\ddots}{}{}{}{\lambda_q} \threevec{\v_1^\top}{\vdots}{\v_q^\top}
\eneqn
where the $\v$'s are the $q \times 1$ eigenvectors of $\widetilde{\phiX}^\top\widetilde{\phiX}$ joined column-wise into the matrix $\bv V$, and the $\lambda$'s are the corresponding eigenvalues (which are non-negative) that form the diagonal of the matrix $\bv \Lambda$. By convention, the eigenvalues are sorted in decreasing order, and their eigenvectors follow suit when packed into $\bv V$. Recall that as a consequence of eigendecomposition, each of the eigenvectors are mutually orthogonal.\footnote
{
$\bv V$ is $q \times N$, so that each row represents an expansion term and each column represents an observation. $\bv \Lambda$ is $N \times N$. The covariance matrix can be expressed as a simple function of its eigenvalues and eigenvectors. 
} 

Without loss of generality, we consider the orthonormal set of eigenvectors $\bv V$, which are obtained by rescaling each eigenvector by the reciprocal of its norm ${\v_k} \leftarrow (\sum_{i=1}^q v_{k,i}^2)^{-1/2} \v_k$).

Note that with $N$ data points, the matrix $\widetilde{\phiX}^\top\widetilde{\phiX}$ can be at most rank $N$ (it can be less than $N$, but for simplicity, we assume it is exactly rank $N$). Thus, when $q > N$, we only need consider the first $N$ eigenvectors because the eigenvalues of the remaining $q-N$ eigenvalues will all be 0. It follows that Equation~\ref{eq"blsh} can be rewritten as

\bneqn
\label{eq"blsh2}
\widetilde{\phiX}^\top \widetilde{\phiX} = [{\v}_1 \ldots {\v}_N] \threebythreemat{{\lambda}_1}{}{}{}{\ddots}{}{}{}{{\lambda}_N} \threevec{{\v}_1^\top}{\vdots}{{\v}_N^\top} = \V \bv \Lambda \V^\top.
\eneqn

In standard principle components regression, it is conventional to compute the eigenvalues normalized by their sum, $\lambda'_k = \lambda_k / \sum_{j=1}^N \lambda_j$. This results in the convenient interpretation that each of the $\lambda'_k$'s represent the percentage of variation explained by the $k$th most important dimension. Because the $\lambda$'s are sorted from high to low, the first eigenvector $\v_1$ represents the \qu{most important} dimension, the second eigenvector $\v_2$ represent the second most important dimension, and so on. One can decide from the cumulative sum of the $\lambda'_k$'s how many eigenvectors (and thereby PCs) should be used in the subsequent analysis. Earlier, we used $\rho$ to denote this cumulative sum. A $\rho$ of 90\% means that the number of PCs retained for later use ``accounted for'' 90\% of the variance of the covariance matrix.

However, in KPCR for logistic regression, instead of selecting a number of PCs directly using solely the value of $\rho$, one proceeds in a three step process that begins by trying to match the validation data cost ratio as closely as possible the desired cost ratio (see Section 3.6 and Figure~\ref{fig:t}). In KPCR for a numerical response variable, there is no cost ratio to approximate and selection depends on the lowest out-of-sample squared error in the validation data (not shown in this paper).

After deciding to employ the first $r$ eigenvectors, one must transform each observation in the expanded space ${\Phi(\x_1)}, \ldots, {\Phi(\x_n)}$. All were $1 \times q$ vectors in $\x_1', \ldots, \x_n'$ after being expanded. With the application of PCA, the dimension can be reduced to $1 \times r$ vectors by a transformation that eliminates the minor $q-r$ dimensions. How does one accomplish this transformation? Some straightforward linear algebra shows that orthogonal projection onto a vector $\v_k$ is given by

\bneqn
\label{eq:rotation2}
\X_k' = \phiX \v_k.
\eneqn
The notation $\X_k'$ (without the \qu{$\Phi(\cdot)$}) is used to indicate that this is the $k$th column of the new regressor matrix, or $k$th PC arrived at through projection of $\phiX$ onto the subset of eigenvectors of the expanded basis. The new regressor matrix can be used in linear model in the same way that the original regressor matrix $\X$ was used.\footnote
{
After taking the transposes, $\v_k^\top$ is $1 \times q$, and $\phiX^\top$ is $q \times N$. The projected values for the $k$th column of the new regressor matrix are the $N$ linear combinations of expansion terms of $\Phi(\textbf{X})$, each weighted the $k$th eigenvector values. They have some of the look and feel of regression fitted values. 
}

However, this procedure only works if one can calculate $\V$. That creates a serious problem because one can only calculate $\V$ if $\phiX$ is known, and the transformation $\Phi$ is unknown. Fortunately, the ``kernel trick'' permits recovery of $\X'$ without knowledge of $\Phi$. 

\subsection*{The Singular Value Decomposition and the Kernel Trick}

Consider the usual singular value decomposition (SVD)  that is a tripartite decomposition valid for any matrix. Applying the SVD, our expanded bases matrix can be decomposed into $\widetilde{\phiX} = \U \bv \Lambda^{1/2} \V^\top$, where $\V$ is as above; it is the matrix of the column-wise eigenvectors of $\widetilde{\phiX}^\top \widetilde{\phiX}$ sorted in decreasing eigenvalue order and becomes size $q \times N$ after dropping the dimensions associated with an eigenvalue of zero. $\bv \Lambda$ also is as above, implying that the middle matrix in the SVD is diagonal and is composed of the square roots of the eigenvalues. $\U$ is the matrix of the eigenvectors of $\widetilde{\phiX} \widetilde{\phiX}^\top$ likewise sorted in decreasing eigenvalue order and is $N \times N$. 

In Section 2.2 we named $\K = \phiX \phiX^\top$ the \qu{kernel matrix.} Here, we name $\widetilde{\K} = \widetilde{\phiX} \widetilde{\phiX}^\top$ the ``centered kernel matrix.''\footnote
{
This is achieved implicitly by centering the kernel matrix $\bv K$ to $\bar{\bv K}=\K - \bv 1_N \K - \K \bv 1_N + \bv1_N \K \bv 1_n$ where $\bv1_N$ is an $N\times N$ matrix of elements $1/N$. the centering is need so that different means across predictors do not dominate the results.

}
Simple linear algebra shows that they are related via

\bneqn\label{eq:centeredkernelmatrix}
\widetilde{\K} = \K - \frac{1}{N} \bv J_N \K - \frac{1}{N} \K \bv J_N + \frac{1}{N^2} \bv J_N \K \bv J_n,
\eneqn
where $\bv{J}_N$ represents a $N \times N$ matrix with all entries being 1. Another property of SVD is that $\widetilde{\K}$ has the same eigenvalues as $\bv \Lambda$.

Additionally, SVD makes $\V$ an orthonormal basis for the rowspace of $\widetilde{\phiX}$ and $\U$ an orthonormal basis for the column space of $\widetilde{\phiX}$. They are related via

\bneqn
\label{eq:v_def}
\widetilde{\phiX} \v_k = \sqrt{\lambda_k} \u_k \quad \text{and} \quad \widetilde{\phiX}^\top \u_k = \sqrt{\lambda_k} \v_k.
\eneqn

Thus, an arbitrary eigenvector $\v_k$ can be written as

\bneqn
\label{eq:v_def2}
\v_k^\top = \oneoversqrt{\lambda_k} \u^\top_k \widetilde{\phiX}.
\eneqn

Finally, one has the solution for not having $\V$. One can project $\Phi(\X)$ onto the $s$-dimensional subset of $\bv V$ \textit{despite being unable to construct $\bv V$ explicitly}. This is the key step in kernel regression and is somewhat counterintuitive.

The projected $\X_k'$ of $\phiX$ onto the $k$th eigenvector $\v_k$ is given by Equation~\ref{eq:rotation2}. We now substitute Equation~\ref{eq:v_def2}, which we learned from the SVD, into the projection formula to arrive at

\bneqn 
\label{eq:trick}
\X_k'^\top = \v_k^\top \phiX^\top = \parens{\oneoversqrt{\lambda_k} \u^\top_k \widetilde{\phiX}} \widetilde{\phiX}^\top = \oneoversqrt{\lambda_k} \u^\top_k \widetilde{\K}.
\eneqn 
Equation~\ref{eq:trick} employs the kernel trick (the equivalence of the outer product in the centered expanded bases with the centered kernel matrix).

\subsection*{A Summary}

In summary, one considers the eigenvalues $\bv \Lambda$, which are calculated via the eigendecomposition of $\widetilde{\K}$ (because it shares the same eigenvalues as $\widetilde{\phiX}^\top \widetilde{\phiX}$). One then picks the first $r < N$ eigenvectors to form a subspace that explains a large enough percentage of the variance in $\C$. Then $\phiX$ is rotated onto this lower dimensional space to obtain the new regressor matrix $\X'$. Taking the transpose of Equation~\ref{eq:trick} and absorbing the $1/\sqrt{\lambda_k}$ into $\U$\footnote{Scaling predictors will not affect fitted values in a linear model. Moreover, the columns of $\X'$ are generally uninterpretable and are not an inferential target.}, one arrives at the simple

\bneqn
\label{eq:rotationfinal}
\X' = \widetilde{\K} \U_{1\ldots r},
\eneqn
where the $\U_{1\ldots r}$ denotes the first $r$ columns of the full $\U$ matrix.\footnote
{
$\widetilde{\K}$ is $N \times N$ and $\U_{1\ldots r}$ is $N \times r$.
} 

\subsection*{Forecasting for New Cases}

Fitted values then follow as usual via ordinary least squares or logistic regression, and with each new $\x^*$ for which one wishes to obtain a forecast. But one must first recapitulate the steps that that transform the original regressors into the principle components used in the logistic regression. That is, $\x^*$ must transformed  into $\Phi(\x^*)$ then rotated onto the selected $\v_1 \ldots \v_r$ chosen during the modeling phase. Following Equation~\ref{eq:rotation2}, one obtains $\x'^{*\top}_k = \v_k^\top \Phi(\x^*)^\top$. The kernel trick is then used to resolve $\v_k$ in the style of Equation~\ref{eq:v_def2}. Again absorbing the eigenvalue constants into $\U$, transposing as in Equation~\ref{eq:rotationfinal} and generalizing for all $r$ dimensions, the result is
\bneqn
\label{eq:kpca_prediction}
\x'^* = \widetilde{\K}(\x^*, \X) \U_{1\ldots r},
\eneqn
where the function $\widetilde{\K}(\x^*, \X)$ is a function that returns the $1 \times N$ vector of the kernel evaluated between $\x^*$ and all $\x_1, \ldots, \x_N$ and then centered. Following Equation~\ref{eq:centeredkernelmatrix}, it can be shown that

\beqn
\widetilde{\K}(\x^*, \X) = {\K}(\x^*, \X) - \frac{1}{N} \bv 1_N^\top \K - \frac{1}{N} {\K}(\x^*, \X) \bv 1_N \bv 1_N^\top + \parens{\frac{1}{N^2} \bv 1_N^\top \K \bv 1_N} \bv 1_n^\top
\eeqn
where $\bv 1_N$ is the $n \times 1$ vector of all 1's.
Finally, to get a prediction for $\x^*$, we take the rotated $\x'^* $ and use the slope estimates from the generalized linear model in the usual fashion. Hence, computing a predicted value requires not only the estimated regression coefficients, but also the original data $\x_1, \ldots, \x_N$ and the kernel function as well. One cannot simply drop an $\x^*$ into the estimated logistic regression equation. 

\pagebreak
\section*{References}
\begin{description}
\item
Arnold Foundation (2013) ``Developing a National Model for Pretrial Risk Assessment.'' Research Summary from the Laura and John Arnold Foundation, www.arnoldfoundation.org.
\item
Berk, R.A., (2012) \textit{Criminal Justice Forecasts of Risk: A Machine Learning Approach}. New York: Springer.
\item
Berk, R.A., Brown, L., and L. Zhao (2010) ``Statistical Inference After Model Selection.'' \textit{Journal of Quantitative Criminology} 26(2): 217--236.
\item
Berk, R.A., Brown, L., Buja, A., Zhang, K., and Zhao, L. (2013) ``Valid Post-Selection Inference.'' \textit{Annals of Statistics} 41(2): 401--1053.
\item
Berk, R.A., Brown, L., Buja, A., George, E., Pitkin, E., Zhang, K., and Zhao, L. (2014) ``Misspecified Mean Function Regression: Making Good Use of Regression Models That Are Wrong.'' \textit{Sociological Methods and Research}, OnlineFirst. 
\item
Berk, R.A., and Bleich, J. (2013) ``Statistical Procedures for Forecasting Criminal Behavior: A Comparative Assessment.'' \textit{Journal of Criminology and Public Policy} 12(3): 513--544.
\item
Bishop, C.M. (2006) \textit{Pattern Recognition and Machine Learning}. New York: Springer.
\item
Borden, H.G. (1928) ``Factors Predicting Parole Success.'' \textit{ Journal of the American Institute of Criminal Law and Criminology} 19: 328--336.
\item
Bornstein, B.H., Tomkins, A.J., Neeley, E.M., Herian, M.N., and Hamm, J.A. (2013) ``Reducing Courts’ Failure-to-Appear Rate by Written Reminders.'' \textit{Psychology, Public Policyand Law} 19 (1): 70--80.
\item
Brieman, L. (2001) ``Random Forests.'' \textit{Machine Learning}, 45: 5--32.
\item
Buja, A, Berk, R., Brown, L., George, E., Pitkin. E., Traskin, M., Zhang, K., and Zhao. L. (2104) ``A Conspiracy of Random Predictors and Model Violations against Classical Inference in Regression.'' \textit{stat.ME}, arXIV:1404.1578v1.
\item
Burgess, E.~M. (1928) ``Factors Determining Success or Failure on Parole.'' In A.~A. Bruce, A.~J. Harno, E.~.W Burgess, \& E.~W., Landesco (eds.) \textit{The Working of the Indeterminate Sentence Law and the Parole System in Illinois} (pp. 205--249). Springfield, Illinois, State Board of Parole.
\item
Bushway, S.D. (2013) ``Is There Any Logic to Using Logit: Find the Right Tool for the Increasingly Important Job of Risk Prediction.'' \textit{Criminology and Public Policy} 12(3): 563--567.
\item
Chipman, H,A., George, E.I., and McCulloch, R.E. (2010) ``BART: Bayesian Additive Regression Trees.'' \textit{Annals of Applied Statistics} 4(1): 266--298.
\item
Cule, E., and De Iorio, M. (2012) ``A Semi-Automatic method to Guide the Choice of Ridge Parameter in Ridge Regression.'' \textit{Annals of Applied Statistics}, working paper, Department of Epidemiology and Biostatistics, School of Public Health, Imperial College London.
\item
Cule, E., and De Iorio, M. (2013) ``Ridge Regression in Prediction Problems: Automatic Choice of the Ridge Parameter.'' \textbf{Genetic Epidemiology} 37(7): 704--714.
\item
Duvenaud, D., Lloyd, J.R., Grosse, R., Tenenbaum, J.B., and Ghahramani, Z. (2013) ``Structure Discovery in Nonparametric Regression through Compositional Kernel Search.'' \textit{Proceedings of the 30th International Conference on Machine Learning}, Atlanta, Georgia, USA.
\item
Faraway, J.J. (2014) ``Does Data Splitting Improve Prediction?" arXiv:1301.2983v2.
\item
Farrington, D.~P. and Tarling, R. (2003) \textit{Prediction in Criminology.} Albany: SUNY Press.
\item
Friedman, J.H. (2002) ``Stochastic Gradient Boosting.'' \textit{Computational Statistics and Data Analysis} 38: 
367--378.
\item
Goldkamp, J. S., and White, M. D. (2006). ``Restoring Accountability in Pretrial Release: The Philadelphia Pretrial Release Supervision Experiments.'' \textit{Journal of Experimental Criminology} 2, 143--181. 
\item
Gottfredson, S.~D., and Moriarty, L.~J. (2006) ``Statistical Risk Assessment: Old Problems and New Applications.'' \textit{Crime \& Delinquency}52(1): 178--200.
\item
Grosse, R.B., Salakhutdinov, R., Freeman, W.T., and Tennebaum, J.B. (2012) ``Exploiting Compositionality to Explore a Large Space of Model Structures.'' 30th Conference on Uncertainty in Artificial Intelligence, Quebec City, Quebec, Canada.
\item
Hastie, T.J. and Tibshirani, R.J. (1990). \textit{Generalized Additive Models}. New York: Chapman \& Hall/CRC. 
\item
Hastie, T., Tibshirani, R., and Friedman, J. (2009) \textit{Elements of Statistical Learning: Data Mining. Inference, and Prediction}, second edition. New York: Springer.
\item
Hoerl, A.E., Kennard, R.W., and Baldwin, KF. (1975) ``Ridge Regression: Some Simulations. \textit{Communications in Statistocs - Theory and Methods} 4:105 -- 123.
\item
Leeb, H. and B.M. P\"{o}tscher (2005) ``Model Selection and Inference: Facts and Fiction.'' \textit{Econometric Theory} 21: 21--59.
\item
Leeb, H., B.M. P\"{o}tscher (2006) ``Can One Estimate the Conditional Distribution of Post-Model-Selection Estimators?'' \textit{The Annals of Statistics} 34(5): 2554--2591.
\item
Le Cressie, S., and Van Houwelingen (1992) ``Ridge Estimators ro Logistic Regression.'' \textit{Journal of Applied Statistics} 41 (1): 191--201.
\item
Lockhard, R., Taylor, J., Tibshirani, R., and Tibshirani, R. (2014) ``A Significance Test for the Lasso.'' (with discussion) \textit{Annals of Statistics}, forthcoming.
\item
McElroy, J.E. (2011) ``Introduction to the Manhattan Bail Project.'' \textit{Federal Sentencing Reporter} 24(1): 8--9.
\item
Mease, D., Wyner, A.J., and Buja, A. (2008) ``Boosted Classification Trees and Class Probability/Quantile Estimation.'' \textit{Journal of Machine Learning Research} 8: 409--439.
\item
Qin, Z., B. Huang, B., Chandramouli, S.S, He, J., and Kumar, S. (2011) ``Large-scale Sparse Kernel Logistic Regression - with a Comparative Study on Optimization Algorithms. Spotlight Oral and Poster session at 6th Annual NYAS Machine Learning Symposium. 
\item
Reiss, A.J. (1951) ``The Accuracy, Efficiency, and Validity of a Prediction Instrument.'' \textit{American Journal of Sociology}
56: 552--561.
\item
Ridgeway, G. (2013a) ``The Pitfalls of Prediction.'' \textit{NIJ Journal} 271.
\item
Ridgeway, G. (2013b) ``Linking Prediction to Prevention.'' \textit{Criminology and Public Policy} 12(3): 545--562.
\item
Searle, S.R. (1982) \textit{Matrix Algebra Useful for Statistics} New York: Wiley.
\item
Schaefer, R., Roi, L., and Wolfe, R. (1984) A Ridge Logistic Estimator. \textit{Communications in Statistics -- Theory and Methods} 13: 99–113.
\item
Sch\"{o}lkopf, B., Herbrich,R.,and Smola, A. J. (2001). ``A Generalized Representer Theorem.'' \textit{Computational Learning Theory. Lecture Notes in Computer Science} 2111: 416 -- 426.
\item
VanNostrand, M., and Keebler, G. (2009). \textit{Pretrial Risk Assessment in the FederalCoourt.} Washington, DC: Office of the Federal Detention Trustee, U.S. Department of Justice.
\item
Vapnick, V. (1998) \textit{Statistical Learning Theory}. New York; Wiley.
\item
Voorman, A., Shokaie, A., and Witten, D. (2014) ``Inference in High Dimensions with the Penalized Score Test.'' posted on arXiv: 1401.2678v1
\item
Wahba, G., Wang, Y., Gu, C., Klein, R., and Klein, B. (1994) ``Smoothing Spline ANOVA for Exponential Families, with Application to the Wisconsin Epidemiological Study of Diabetic Retinopathy.'' \textit{The Annals of Statistics} 33(6): 1865--1895.
\item
White, H. (1980) ``Using Least Squares to Approximate Unknown Regression Functions.'' \textit{International Economic Review} 21(1): 149--170.
\item
White, H. (1982) ``Maximum Likelihood Estimation of Misspecified Models.''\textit{Econometrica} 50(1): 1--25.
\item
Zhu, J., and Hastie, T. (2005) ``Kernel Logistic Regression and The Import Vector Machine.'' \textit{Journal of Computational and Graphical Statistics} 14: 185--205.

\end{description}

\end{document}